\documentclass[twocolumn,prl,superscriptaddress]{revtex4-1}

\usepackage[english]{babel}
\usepackage{boldline,multirow,xcolor,colortbl}
\usepackage[ansinew]{inputenc}
\usepackage{times}
\usepackage{graphicx}
\usepackage{graphics}
\usepackage{amsmath}
\usepackage{amsfonts}
\usepackage{amssymb}
\usepackage{dsfont}
\usepackage{epstopdf}
\usepackage{makeidx}
\usepackage{subfigure}
\usepackage{color}
\usepackage{pgf}
\usepackage{bm}
\usepackage{tikz} 
\usepackage[normalem]{ulem}

\newcommand{\abs}[1]{\vert #1\vert}

\makeatletter 
\renewcommand{\thefigure}{\@arabic\c@figure}
\makeatother

\newcounter{defcounter}
\newenvironment{myeqnarray}{%
\addtocounter{equation}{-1}
\refstepcounter{defcounter}

\begin{eqnarray}}
{\end{eqnarray}}

\begin{document}

\title{Casimir Force Phase Transitions in the Graphene Family}

\author{Pablo Rodriguez-Lopez}
\affiliation{Department of Physics, University of South Florida, Tampa FL, 33620, USA}

\author{Wilton J. M. Kort-Kamp}
\affiliation{Center for Nonlinear Studies, MS B258, Los Alamos National Laboratory, Los Alamos, NM 87545, USA}
\affiliation {Theoretical Division, MS B213, Los Alamos National Laboratory, Los Alamos, NM 87545, USA}

\author{Diego A. R. Dalvit}
\affiliation {Theoretical Division, MS B213, Los Alamos National Laboratory, Los Alamos, NM 87545, USA}

\author{Lilia M. Woods} 
\affiliation{Department of Physics, University of South Florida, Tampa FL, 33620, USA}


\begin{abstract}
\bf{The Casimir force is a universal interaction induced by electromagnetic quantum fluctuations between any types of objects. The expansion of the graphene family by adding silicene, germanene, and stanene, 2D allotropes of Si, Ge, and Sn, lends itself as a platform to probe Dirac-like physics in honeycomb staggered systems in such a ubiquitous interaction. 
We discover Casimir force phase transitions between these staggered 2D materials
induced by the complex interplay between Dirac physics, spin-orbit coupling, and externally applied fields. 
In particular, we find that the interaction energy experiences different power law distance decays, magnitudes, and dependences on characteristic physical constants. Furthermore,
due to the topological properties of these materials, repulsive and quantized Casimir interactions become possible. }
\end{abstract}

\maketitle

\hspace{-12pt} Interactions originating from electromagnetic quantum fluctuations are universal as they exist between objects regardless of their  specific properties or boundary conditions. These ubiquitous interactions lead to the well-known van der Waals (vdW) 
force \cite{Parsegian_BookVdW} when the exchange of electromagnetic fluctuations can be considered instantaneous, and to the Casimir and Casimir-Polder forces 
when the distances between the objects are large and the finite speed of light is important \cite{Casimir_Polder, Casimir_PMPlates}. 
Although these interactions are typically weak, they have appreciable effects at nano- and micro-meter separations. For example, adhesion, stiction, wetting, and stability of materials composed of chemically inert constituents occur due to vdW/Casimir interactions \cite{RPM_Galina_Klimchitskaya,Rodriguez-2011,Dalvit:book,Woods-RevModPhys}. 
The discovery of systems with reduced dimensions and physics different from the one of standard 3D dielectrics, metals, and semiconductors  has given a new impetus to the field of vdW/Casimir phenomena. Specifically, systems involving 
graphene \cite{Novoselov666_Graphene} have a strong dependence on temperature and doping in their vdW/Casimir interactions
\cite{PhysRevLett.96.073201,PhysRevB.80.245424,PhysRevB.82.155459,PhysRevB.84.155407,Quantized_Casimir_Force,PhysRevB.89.125407,VGobre-2014}. 
Experimental measurements have demonstrated that the vdW force between substrates is almost completely screened when one is covered by graphene \cite{Tsoi-2014}, while temperature effects in graphene-based Casimir interactions have also been reported \cite{Mohideen-2013}. 

Recently the graphene family has expanded. Silicon, germanium, and tin, being in the same column of the periodic table as carbon, also have stable 2D layers \cite{PhysRevLett.108.155501, Davila-2014, RIS_0}. Unlike the planar sp$^{2}$ bonded graphene, silicene, germanene, and stanene have spatial buckling between the two sublattices caused by their stronger sp$^{3}$ bonding. These newer members of the 2D graphene family exhibit non-trivial topological insulator features. The application of external fields together with the inherently strong spin-orbit coupling can be used as effective ``knobs'' for various Hall transitions \cite{PhysRevLett.107.076802,Ezawa-2012,Ezawa-PhysRevB.86.161407,Ezawa2013,PhysRevLett.111.136804,Houssa2016,PhysRevB.86.195405,PhysRevLett.110.197402,PhysRevB.87.235426,PhysRevB.88.045442}.
Furthermore, vertically stacking of different 2D materials held by vdW interactions is emerging as a new scientific direction, where desired properties by design can be achieved \cite{VdW_heterostructures, Lin-vdWsolids-2014}. Recent  studies have shown that the vdW interactions affect the electronic and phonon properties of such vdW heterostructures \cite{Terrones-2014, Le-2016}, which is especially relevant for their transport and optical applications. 

In this paper, we study the physics of Casimir interactions in the graphene family, which serve as a platform for probing low-energy Dirac-like physics in systems that can experience different Hall transitions. 
We find that phase transitions between the various electronic phases in these materials, attained by means of externally applied circularly polarized lasers and/or static electric fields, strongly impact fluctuation-induced phenomena. Novel distance scaling laws, abrupt magnitude changes, force quantization and repulsion, are all manifestations of Casimir force phase transitions occurring in these 2D staggered materials. \\

%%%%%%%%%%%%%%

\hspace{-12pt} {\bf Results}

\begin{figure}
\centering
\includegraphics[width=1.0\linewidth]{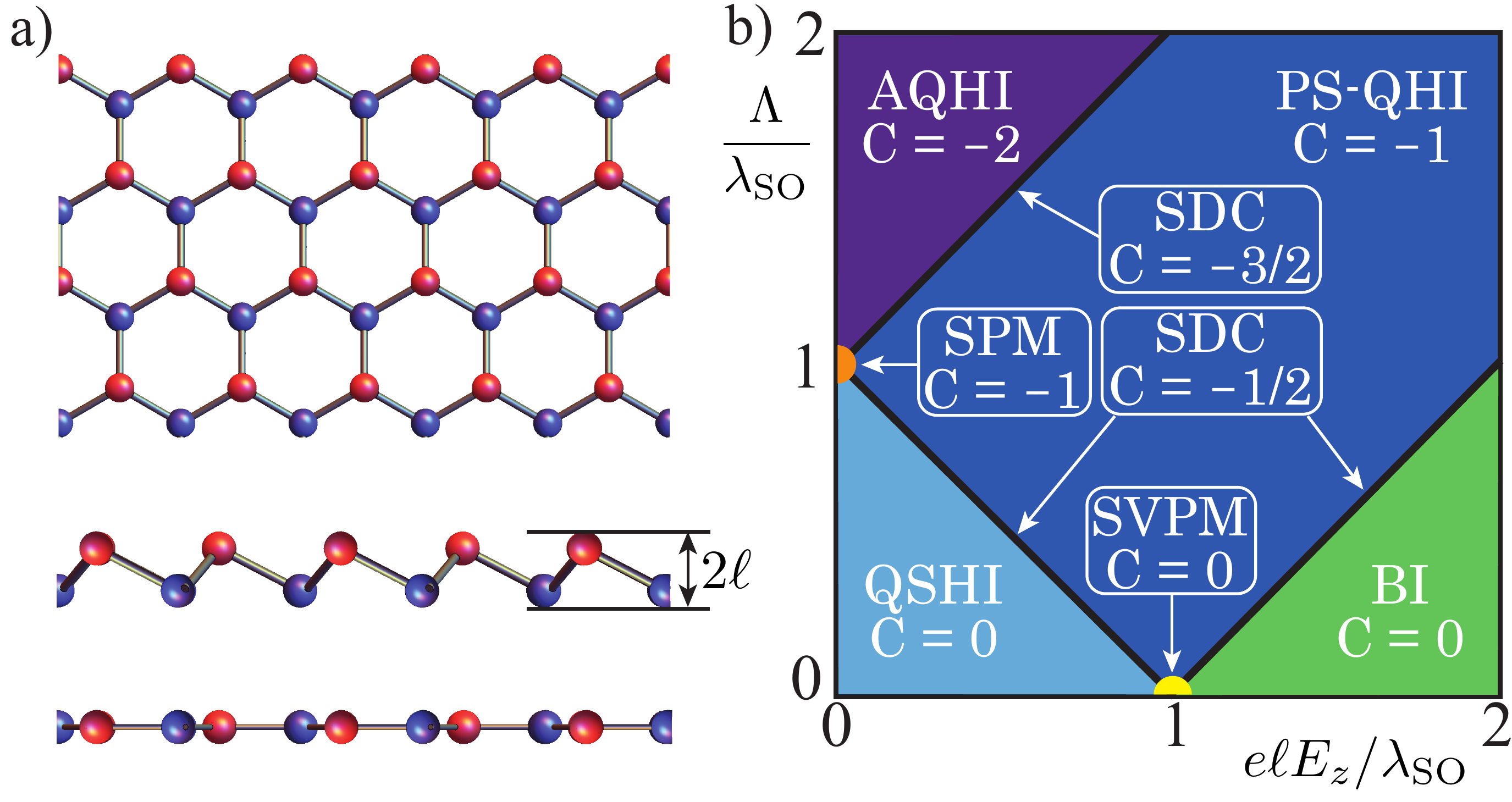}
\caption{
{\bf Phase diagram of the graphene family.}
(a) Top view of the hexagonal lattice structure of the graphene family. The red and blue colors represent the two inequivalent atoms in the structure. While graphene has planar atomic configuration (side view shown), the graphene-family materials (silicene, germanene and stanene) have a finite staggering $2 \ell$ between the two sublattices. 
(b) First quadrant of the phase diagram of the graphene-family materials in the $(e \ell E_{z},\Lambda)$ plane in units of $\lambda_{\text{SO}}$ ~\cite{Ezawa2013}. The distinct electronic phases (acronyms are defined in the main text) are characterized by the Chern number $C$. 
}
\label{fig:Fig1}
\end{figure}

\hspace{-12pt} {\bf Electro-optical response of the 2D graphene family.}
Silicene, germanene, and stanene have layered honeycomb structure similar to graphene, but
the two inequivalent atoms in the unit cell are arranged in staggered layers characterized by a finite buckling $2\ell$, as shown in Fig. \ref{fig:Fig1}(a)
\cite{PhysRevB.84.195430,PhysRevLett.107.076802,PhysRevLett.111.136804}. 
In graphene artificial efforts are needed to modify the carrier mass and induce spin orbit coupling (SOC) \cite{Floquet_CI_Graphene, PhysRevLett.112.156801}. However, thanks to the buckling and heavier constituent atoms, such properties are already intrinsic to silicene, germanene, and stanene. The low energy band structure can be determined from a Dirac-like Hamiltonian, obtained from a nearest neighbor tight binding model, which also includes an external electric field $E_z$  perpendicular to the 2D plane of the material and irradiated circularly polarized light 
\cite{Ezawa2013,Ezawa-2015}
\begin{eqnarray}\label{Low_Energy_Hamiltonian}
H_{s}^{\eta} & = & \hbar v_{\text{F}}\left( \eta k_{x}\tau_{x} + k_{y}\tau_{y} \right) + \Delta_{s}^{\eta}\tau_{z} - \mu\tau_{0},
\end{eqnarray}
\begin{eqnarray}\label{Mass_Parameter}
\Delta_{s}^{\eta} & = & \eta s\lambda_{\text{SO}} - e  \ell E_{z} - \eta\Lambda.
\end{eqnarray}
Here, $\tau_{i}$ are the Pauli matrices for the sublattice pseudospin index $\eta=\pm 1$, $\tau_{0}$ is the identity matrix, and the spin index $s=\pm 1$ denotes the eigenvalues of the Pauli spin matrix $\sigma_{z}$. Also, 
$e$ is the electron charge,
$\mu$ is the chemical potential and $v_{\text{F}} = \sqrt{3}a t/2 \hbar$ is the Fermi velocity, where 
$a$ is the lattice constant
($a^{\text{Gra}}=2.46 \text{\AA}$, $a^{\text{Sil}}=3.86 \text{\AA}$, $a^{\text{Ger}}=4.02 \text{\AA}$, and $a^{\text{Stan}}=4.7 \text{\AA}$), and
$t$ is the nearest-neighbor coupling ($t^{\text{Gra}}=2.8$ eV, $t^{\text{Sil}}=1.6$ eV, $t^{\text{Ger}}=1.3$ eV, and $t^{\text{Stan}}=1.3$ eV).
For graphene, $\ell^{\text{Gra}}=\lambda^{\text{Gra}}_{\text{SO}}=0$ and for the other materials, $\ell$ has values that are of similar order ($\ell^{\text{Sil}}=0.23\,\text{\AA}$, $\ell^{\text{Ger}}=0.33\,\text{\AA}$ and $\ell^{\text{Stan}}=0.40\,\text{\AA}$), but $\lambda_{\text{SO}}$ can vary by orders of magnitude ($\lambda^{\text{Sil}}_{\text{SO}}=3.9$ meV, $\lambda^{\text{Ger}}_{\text{SO}}=43$ meV and $\lambda^{\text{Stan}}_{\text{SO}}=100$ meV) ~\cite{Ezawa-2015}. The components of the 2D wave vector in Eq. \ref{Low_Energy_Hamiltonian} are denoted as $k_{x,y}$ and $\Delta_{s}^{\eta}$ is the Dirac mass at the $K_{\eta}$ points for each spin index $s$, characterized by the eigenenergy 
$E_s^{\eta}= \pm\sqrt{ \hbar^{2} v_{\text{F}}^{2} k^{2} + ({\Delta_{s}^{\eta}})^{2} } - \mu$. 

The mass parameter in Eq. \ref{Mass_Parameter} depends on the strength of the SOC and the spin and valley degrees of freedom of the  carriers. It can be further controlled by $E_{z}$, which generates an electrostatic potential $2\ell E_{z}$ between the two different atoms in the unit cell. 
Other types of SOC originating from Rashba physics, such as the Rashba SOC associated with the next-nearest neighbor hopping and the Rashba SOC associated with the nearest neighbor hopping induced by $E_{z}$, are neglected here due to their small effects as compared to $\lambda_{\text{SO}}$  \cite{Ezawa-PhysRevB.86.161407}. The properties of all 2D materials can also be modified by irradiating circularly polarized light, with the electromagnetic vector potential  given by 
${\bf A}(t)=A_0 (\cos(\omega_{0}t), \pm \sin(\omega_{0}t), 0)$, where $A_0$ is an amplitude and $\omega_{0}$ is the frequency of the applied light with $+(-)$ specifying left (right) circular polarization.   In the limit 
$(e a A_0/\hbar)^2 \ll 1$, and using a low-energy Hamiltonian approach, this results in a contribution to the Dirac mass gap given by $\Lambda = \pm (e v_{\text{F}} A_0)^2 /c^2 \hbar \omega_0$ (we use cgs electromagnetic units) \cite{Ezawa2013}, as shown in Eq. \ref{Mass_Parameter}. 
We should note that the light field may also cause additional coupling between the energy bands \cite{Oka-2009}, which can open gaps in the band structure typically at energies around $n \hbar \omega_{0}/2$ ($n=\pm 1, \pm 2$, etc). 
 Hence, the Hamiltonian in Eq.(1) is valid as long as $| E_{s}^{\eta}| < \hbar \omega_0/2$.

The staggered 2D layers exhibit several electronic phases \cite{Ezawa-2012,Ezawa2013} resulting from changes in $\Delta_{s}^{\eta}$ induced by $E_{z}$ and/or $\Lambda$ (see Fig. \ref{fig:Fig1}(b)).
At $E_z=\Lambda=0$, the 2D layer can be characterized as a Quantum Spin Hall Insulator (QSHI). Fixing $\Lambda=0$ and increasing $E_z$, it remains in the QSHI phase until the critical electric field $E_{z, \text{cr}} = \lambda_{\text{SO}}/e \ell$ is reached. At this point, two Dirac cones are closed ($\Delta_{1}^{1}=\Delta_{-1}^{-1}=0$) and the material becomes a Spin Valley Polarized Semimetal (SVPM). Further increasing the electric field $E_{z}>E_{z,\text{cr}}$, the magnitude of all four $\Delta_{s}^{\eta}$ increases and the 2D layer becomes a regular Band Insulator (BI).
In the case that we fix $E_z=0$ and increase $\Lambda$, the system goes through a phase transition from the QSHI phase to a Spin Polarized Metal (SPM) phase at the critical value $\Lambda_{\text{cr}} = \lambda_{\text{SO}}$, where the energy gap of one of the spins closes. 
For $\Lambda>\Lambda_{\text{cr}}$, the Anomalous Quantum Hall Insulator (AQHI) phase is reached.
When both $E_z$ and $\Lambda$ are non-zero, these materials can have other topological phases \cite{Ezawa2013}. 
For example, the region of the phase diagram in Fig. 1b) where $0\leq e \ell E_{z} + \Lambda < \lambda_{\text{SO}}$ corresponds to a QSHI phase.
Along the line $ e \ell E_{z} + \Lambda =\lambda_{\text{SO}}$ it is possible to have only one Dirac cone closed, the Single Dirac Cone (SDC) phase.
Finally, when the conditions $e \ell E_{z} + \Lambda > \lambda_{\text{SO}}$
and $|e \ell E_{z} - \Lambda| > \lambda_{\text{SO}}$ are simultaneously satisfied, 
the closed gap opens again but with the opposite sign
resulting in a Polarized Spin Quantum Hall Insulator (PS-QHI) state, a combination of the AQHI and QSHI phases. 
For completeness, we briefly describe the other three quadrants of the phase diagram
in Fig. \ref{fig:Fig1}(b). The second quadrant ($E_z<0, \Lambda>0$) is obtained from the first one by taking its mirror replica with respect to the $E_z=0$ axis. The third and fourth quadrants are obtained by taking the mirror replica of the first two with respect to the $\Lambda=0$ axis and inverting the signs of the Chern numbers.

\begin{figure}[t]
\centering
\includegraphics[width=1.0\linewidth]{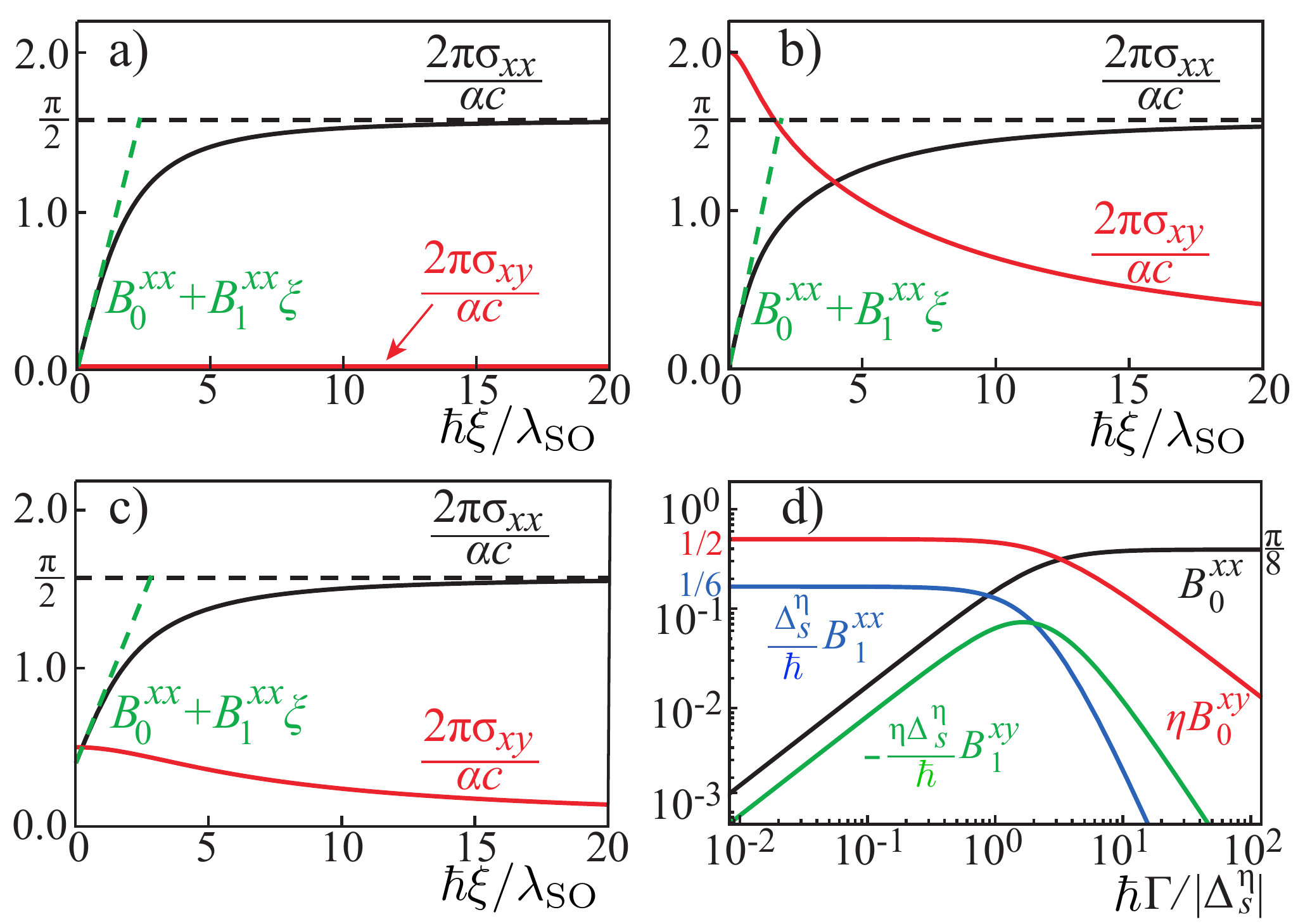}
\caption{
{\bf Zero-temperature longitudinal and Hall conductivities at imaginary frequencies.} The behaviour of 
$\sigma_{xx}(i \xi)$ and $\sigma_{xy}(i \xi)$
for different phases is shown: (a) $E_z=\Lambda=0$ (QSHI phase with $C=0$); (b) $\Lambda/\lambda_{\text{SO}}=-3/2$ and $E_{z}=0$ (AQHI phase with $C=2$); and (c) $e \ell E_z/\lambda_{\text{SO}}=-\Lambda/\lambda_{\text{SO}}=1/2$ (SDC phase with $C=1/2$).  In all cases $\mu = \Gamma = 0$.   
The horizontal black dashed line is $\sigma_{xx}(i\xi\to \infty) = \alpha c/4$.  
The dashed green lines correspond to the low-frequency expansion for the conductivities given in Eqs. (\ref{ConductivityExpansion}). 
For $\mu=0$, $B_{-1}^{xx}$ vanishes identically, while the other coefficients 
$B_0^{xx,xy}$ and $B_1^{xx,xy}$ are shown in panel (d)  as a function of $\hbar \Gamma/|\Delta^{\eta}_s|$. 
}
\label{fig:Fig2}
\end{figure}

The energy band structure has important consequences for the electro-optical response, and in particular for the conductivity tensor at imaginary frequencies, needed for the Casimir force computation (see below). Using the 
standard Kubo formalism \cite{Kubo-I-1957,Kubo-II-1957}, we  obtain the zero-temperature 
dynamical 2D conductivity tensor 
$\sigma_{ij}(i \xi, \Delta_{s}^{\eta})$ of each Dirac cone. Here, $i \xi$ is an imaginary frequency, and $i,j=x,y$ are Cartesian components. For the inter-plate separations and temperatures we study below, effects of spatial dispersion can be neglected \cite{PhysRevB.80.245424,Svetovoy}.
The dynamical conductivity components due to intraband ($\sigma_{ij}^{\text{intra}}$) and interband ($\sigma_{ij}^{\text{inter}}$) transitions are found to be
\begin{equation}
\sigma_{xx}^{\text{intra}}(i\xi, \Delta_{s}^{\eta})=  
\frac{\alpha c}{4\pi}\frac{\mu^{2} - (\Delta_{s}^{\eta})^{2}}{\hbar \Omega|\mu|} \Theta\left( |\mu| - |\Delta_{s}^{\eta}| \right), 
\nonumber
\end{equation}
\begin{equation}
\sigma_{xx}^{\text{inter}}(i\xi, \Delta_{s}^{\eta})=\frac{\alpha c}{4\pi}\frac{(\Delta_{s}^{\eta})^{2}}{\hbar \Omega M} + \frac{\alpha c}{8 \pi }\left[ 1 - \left( \frac{2\Delta_{s}^{\eta}}{\hbar\Omega} \right)^2
\right]
\tan^{-1}\left(\frac{\hbar \Omega}{2 M}\right) ,
\nonumber
\end{equation}
\begin{equation}
\sigma_{xy}^{\text{intra}}(i\xi, \Delta_{s}^{\eta}) = 0 ,
\nonumber
\end{equation}
\begin{equation} 
\sigma_{xy}^{\text{inter}}(i\xi, \Delta_{s}^{\eta})=  \frac{\alpha c}{2\pi}\frac{\eta\Delta_{s}^{\eta}}{\hbar \Omega} \tan^{-1}\left(\frac{\hbar\Omega}{2 M}\right) ,
\label{sigma_Dirac_Cone}
\end{equation}
where $\sigma_{yy}(i\xi, \Delta_{s}^{\eta})=\sigma_{xx}(i\xi, \Delta_{s}^{\eta})$ and $\sigma_{yx}(i\xi, \Delta_{s}^{\eta})=-\sigma_{xy}(i\xi, \Delta_{s}^{\eta})$. 
Here, $\alpha=e^{2}/\hbar c \approx 1/137$ is the fine structure constant, 
$M=\text{max}( |\Delta_{s}^{\eta}|, |\mu|)$, and $\Omega=\xi+\Gamma$, where $\Gamma = 1/2\tau$ with $\tau$ being the relaxation scattering time.
Corresponding expressions for the silicene optical conductivity at real frequencies have already been 
reported in the literature ~\cite{PhysRevB.88.045442,PhysRevB.87.235426}. The dynamical conductivity from all Dirac cones, necessary for the evaluation of the Casimir interaction, is 
$\sigma_{ij}(i\xi) = \sum_{s,\eta=\pm 1}\left[ \sigma_{ij}^{\text{intra}}(i\xi, \Delta_{s}^{\eta}) + \sigma_{ij}^{\text{inter}}(i\xi, \Delta_{s}^{\eta}) \right].$
The various topological phases associated with the Hall effect, displayed in Fig. \ref{fig:Fig1}(b), are captured via the $\eta\Delta_s^\eta$ product in $\sigma_{xy}$. In Figs. \ref{fig:Fig2} (a-c) we show the different elements of the conductivity tensor as a function of imaginary frequency at various points in the phase diagram. 
Finite temperature effects on the optical conductivity can be found in the Supplementary Note 2.\\

%%%%%%%%%%%%%

%%%%%%%%%%%%%

\hspace{-12pt}
{\bf Low frequency optical response.}
Since the Casimir interaction at large separations is determined mainly by the low frequency response \cite{Woods-RevModPhys}, 
understanding the optical conductivity at $i\xi=0$ is particularly important. We first consider the case
$\Lambda=0$. Graphene has neither staggering nor SOC, and hence $\Delta_{s}^{\eta}=0$ for all cones. Using Eqs. \ref{sigma_Dirac_Cone}, one recovers the well known result 
\cite{RevModPhys.81.109} for the graphene universal conductivity $\sigma_{xx}(i\xi) = \alpha c/4$ and $\sigma_{xy}(i\xi) = 0$ in the non-dissipative limit.  For the other members of the graphene family, their conductivity tensors 
can be cast into the perspective of a Chern insulator description (see Fig. \ref{fig:Fig1}(b)), in which the corresponding Chern number is given by $C = \frac{1}{2}\sum'_{s,\eta=\pm 1}\eta \; \text{sign}[\Delta_{s}^{\eta}]$, and captures the topologically non-trivial features of these 2D materials \cite{Ezawa-PhysRevB.86.161407,Ezawa2013}.  The prime in the summation indicates that only terms with $\Delta_{s}^{\eta} \neq 0$ should be included.
Let us now consider the case for $\Delta_{s}^{\eta} \neq 0$ and, as above, restrict ourselves to the dissipationless limit ($\Gamma=0$). When $\abs{\mu} < \abs{\Delta_{s}^{\eta} }$, we find that $\sigma_{xx}(i\xi=0,\Delta_{s}^{\eta}) = 0$ and $\sigma_{xy}(i\xi=0,\Delta_{s}^{\eta}) = \frac{\alpha c}{4\pi}\eta \; \text{sign}[\Delta_{s}^{\eta}]$ for each cone. 
Thus, the total Hall conductivity is $\sigma_{xy}(i\xi=0) = \frac{\alpha c}{2\pi}C$, which explicitly connects with the Chern insulator topological nature of these materials via the particular electronic phase. For example, the $C=0$ for the QSHI phase at $E_z=0$ results in $\sigma_{xy}(i\xi=0)=0$ (Fig. \ref{fig:Fig2}(a)). The $C=2$ AQHI phase at $\Lambda/\lambda_{\text{SO}}=-3/2$ leads to $\sigma_{xy}(i\xi=0) = 2\frac{\alpha c}{2\pi}$ since there are four open Dirac cones and each contributes with the same sign to the Hall conductivity (Fig.~\ref{fig:Fig2}(b)). The $C=1/2$ SDC phase at 
$\ell E_z/\lambda_{\text{SO}}=-\Lambda/\lambda_{\text{SO}}=1/2$ gives $\sigma_{xy}(i\xi=0) = \frac{1}{2}\frac{\alpha c}{2\pi}$ 
since there are three open Dirac cones (Fig.~\ref{fig:Fig2}(c)). 

To gain further insight into the various factors affecting the contribution of each single Dirac cone to the conductivity $\sigma_{ij}(i\xi,\Delta_{s}^{\eta})$ , we perform a low-frequency expansion. Using Eqs. \ref{sigma_Dirac_Cone} one finds 
\begin{eqnarray} 
\sigma_{xx}(i\xi,\Delta_{s}^{\eta}) & = & \frac{\alpha c}{2\pi}
\left[ \frac{B_{-1}^{xx}}{\xi} + B_{0}^{xx} + B_{1}^{xx}\xi + \mathcal{O}(\xi^2) \right], \nonumber \\
\sigma_{xy}(i\xi,\Delta_{s}^{\eta}) & = & \frac{\alpha c}{2\pi}
\left[ B_{0}^{xy} +  B_{1}^{xy}\xi + \mathcal{O}(\xi^2)\right] .
\label{ConductivityExpansion}
\end{eqnarray}
The coefficients $B_{-1}^{xx}$, $B_{0}^{xx,xy}$, and $B_{1}^{xx,xy}$ are a function of the parameters of the 2D material (i.e., $\Delta^{\eta}_s$, $\Gamma$, and $\mu$), and their explicit expressions are given in the Supplementary Note 1. It is interesting to note that each term in Eqs. \ref{ConductivityExpansion} is reminiscent of a particular model dielectric response function. For example, the first term of the longitudinal conductivity behaves as the plasma model for metals with $B_{-1}^{xx}$ specifying the plasma frequency, and it originates entirely from intraband transitions. The Lorentz model for dielectrics is recognized in the third term with $B_{1}^{xx}$ giving the strength of the Lorentz oscillator. $B_{0}^{xx}$ corresponds to a  constant conductivity. On the other hand, 
$B_{0}^{xy}$ captures the Hall effects in the 2D materials, and in the lossless case it can be written as  
$B_{0}^{xy}=C$, which shows the quantized nature of the Hall conductivity via the Chern number. Figs. \ref{fig:Fig2}(a-c) show how the above low-frequency expansion for the longitudinal conductivity compares to the full Kubo expression.

For the case $\mu=0$, $B_{-1}^{xx}$ identically vanishes, and the remaining coefficients are shown in Fig. \ref{fig:Fig2}(d). 
When $\Delta_{s}^{\eta}=0$, 
$B_{0}^{xx}=\pi/8$ and $B_{1}^{xx}= B_{0}^{xy}=B_{1}^{xy}=0$ for all values of the dissipation parameter. 
When $\Delta_{s}^{\eta} \neq 0$, dissipation influences the coefficients. In the limit of small dissipation 
$\hbar \Gamma/|\Delta_s^{\eta}| \ll 1$, 
$B_{0}^{xx} \approx \hbar \Gamma/6 |\Delta_s^{\eta}|$, 
$B_{1}^{xx} \approx \hbar / 6 |\Delta_s^{\eta}|$, 
$B_{0}^{xy} \approx \eta \; \rm{sign}[\Delta_s^{\eta}] / 2$, and
$B_{1}^{xy} \approx - \eta \hbar^2 \Gamma \rm{sign}[\Delta_s^{\eta}] / 12 (\Delta_s^{\eta})^2$.
In the opposite limit $\hbar \Gamma/|\Delta_s^{\eta}| \gg 1$, $B_{0}^{xx}= \pi/8$ and 
all other coefficients tend to zero. \\

%%%%%%%%%%%%%%%%%

\begin{figure}[t]
\centering
\includegraphics[width=1\linewidth]{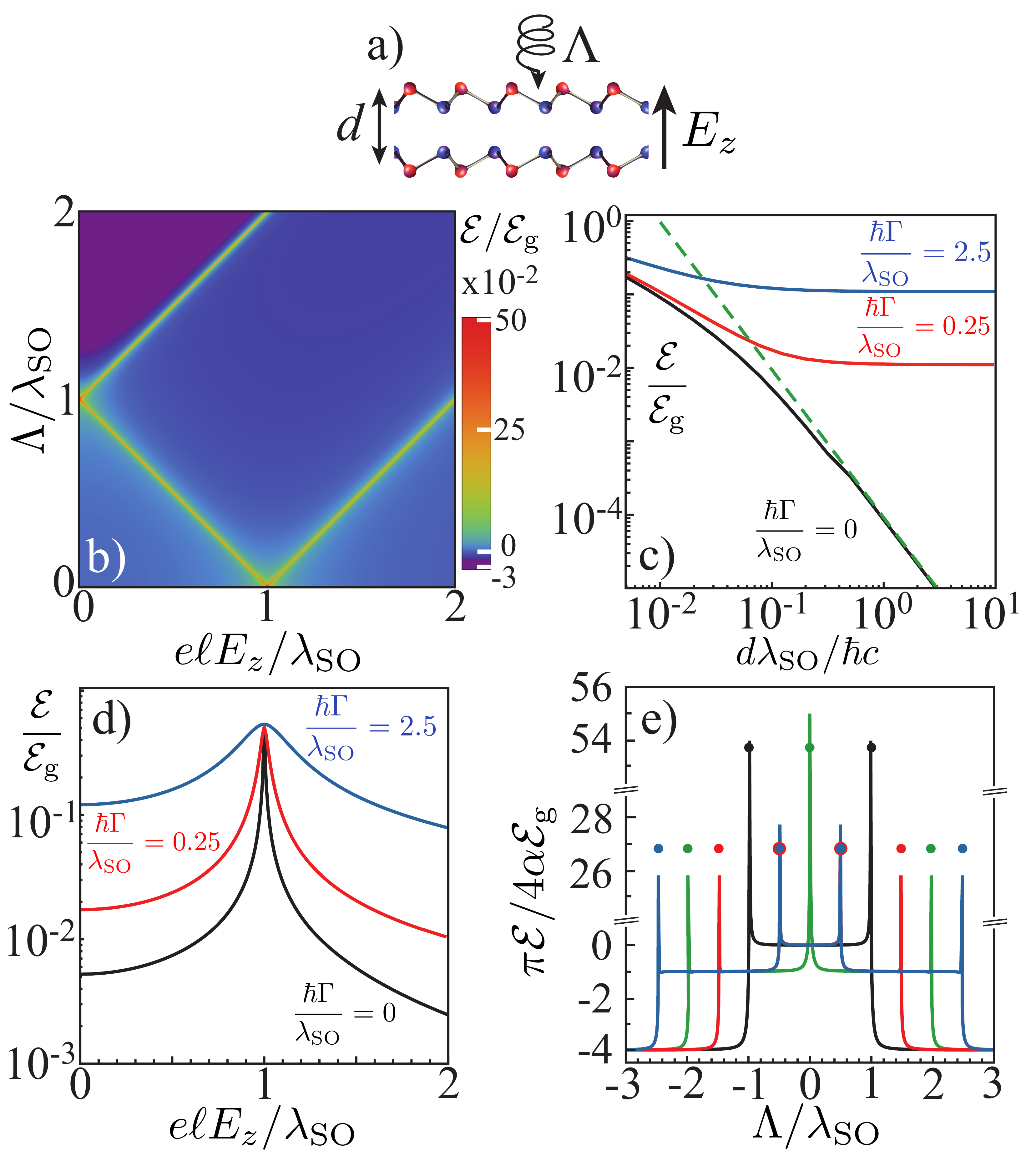}
\caption{
{\bf Zero-temperature Casimir interaction in the graphene family.}
(a) Fabry-P\'{e}rot cavity formed by two layers of the graphene-family materials under externally applied fields.
(b) Casimir energy phase diagram for two dissipationless identical parallel layers for 
$d \lambda_{\text{SO}}/\hbar c=1$.
(c) Distance dependency of the Casimir energy at $\Lambda=E_z=0$ for various values of dissipation.
(d) Effect of the external electric field on the Casimir energy at $\Lambda=0$ for different values of dissipation at a distance
$d \lambda_{\text{SO}}/\hbar c=1$. (e) Quantized and repulsive Casimir energy for two dissipationless identical layers at $d = 10 \hbar c/ \lambda_{\text{SO}}$.
The various curves correspond to different values of the electric field, $e \ell E_{z}/\lambda_{\text{SO}} = \{0, 1/2, 1, 3/2\}$ (black, red, green and blue, respectively).
In the large-distance asymptotics (Table I), the rounded plateaus become abrupt jumps and the interaction energies 
at phase transition boundaries are the dots in-between plateaus. In all plots $\mu=0$.}
\label{fig:Fig3}
\end{figure}

%%%%%%%%%%%%%%%

%%%%%%%%%%%%%%%%%%%

\hspace{-12pt}
{\bf Casimir force phase transitions.}
The Casimir energy per unit area $\mathcal{E}$ and the corresponding Casimir force $F/S=-\partial \mathcal{E}/\partial d$ between two layers of area $S$ of the graphene-family materials separated by a distance $d$ can be calculated using the continuum Lifshitz approach, which applies for separations larger than several times the interatomic distances in the involved objects \cite{Woods-RevModPhys}.
We first discuss the effects of quantum (zero-temperature) fluctuations on the Casimir energy, which in this case is expressed as an integral over complex frequencies $i \xi$ (see Methods).
Beginning with neutral ($\mu=0$) graphene/graphene interaction, the Casimir energy per unit area is found to be
\begin{eqnarray}
\mathcal{E}_{\text{g}} = - \frac{\hbar c\alpha}{32\pi d^{3}},
\label{Eq_Casimir_Graphene}
\end{eqnarray}
and results in Casimir attraction
\cite{PhysRevLett.96.073201,PhysRevB.82.155459,PhysRevB.80.245424}. 
Compared with the Casimir energy for perfect metals $\mathcal{E}_{\text{m}}=- \hbar c\pi^{2} / 720 d^{3}$, it reveals that although the distance dependence is the same, the magnitude of $\mathcal{E}_{\text{g}}$ is much reduced due to the presence of $\alpha$. 

Probing the expanded graphene family optical response by changing $E_z$ and/or $\Lambda$ results in a much richer Casimir interaction picture. The competition between $\sigma_{xx}$ and $\sigma_{xy}$ dominance and the relative contribution of the different coefficients $B_{-1}^{xx}$, $B_{0}^{xx,xy}$, and $B_{1}^{xx,xy}$, result in many different asymptotic scaling laws, significant magnitude changes, force quantization and repulsion. We consider a Fabry-P\'{e}rot cavity formed by two sheets of the graphene family (e.g. Sil/Sil, Ger/Ger, and Stan/Stan) 
(Fig. \ref{fig:Fig3}(a)).
As for graphene, each staggered layer is almost transparent to the incident light  
(transmission coefficient $T \simeq 1- \pi \alpha$), and hence both layers forming the cavity experience irradiation essentially with the same characteristics captured by $\Lambda$. 

The impact of the different phases of the graphene-family materials on the Casimir interaction is shown in 
Fig. \ref{fig:Fig3}(b) for $\mu=\Gamma=0$ and a distance $d=\hbar c/ \lambda_{\text{SO}}$.
The Casimir energy density plot reflects the phase diagram of Fig. \ref{fig:Fig1}(b). Note that for the parameters of the figure, $\mathcal{E}/\mathcal{E}_{\text{g}} < 0$ in most of the AQHI and PS-QHI phases, and will ultimately result in Casimir force repulsion (see discussion below about Fig. \ref{fig:Fig3}(e)). On the other hand, in all other phases $\mathcal{E}/\mathcal{E}_{\text{g}} > 0$ corresponding to attraction. As one approaches 
phase transition boundaries, the Casimir energy significantly increases in magnitude featuring a cusp-like behavior (see black curve in Fig. \ref{fig:Fig3}(d)).  At shorter distances (where non-zero imaginary frequencies become relevant) the energy phase diagram is modified with less defined phase boundaries (not shown).
The dependence of the Casimir interaction energy on separation is shown in Fig. \ref{fig:Fig3}(c) at 
the origin of phase space $\Lambda=E_{z}=0$. The asymptotic result from Table 1 for $\Gamma=0$ is given by the green dashed line in the figure. 
Increasing dissipation in the materials results in a blurring of the phase boundaries and $\mathcal{E}/\mathcal{E}_{\text{g}} > 0$ for all phases at large separations (see Supplementary Note 3).
The behaviour of the phase diagram along the $\Lambda=0$ line for different values of dissipation is shown in Fig. \ref{fig:Fig3}(d).
The Casimir energy for lossless QSHI-QSHI or BI-BI phase combinations changes to that corresponding to the SVPM-SVPM configuration as $E_z$ approaches $E_{z,\text{cr}}$, presenting a cusp-like feature.
At either side of the cusp all Dirac masses are non-zero, while right at the cusp two Dirac cones close. The interaction energy in the SVPM-SVPM configuration has a graphene-like behavior (see Table \ref{Table_GS_SS_Ez_nonzero_A_0}) but, since two rather than four gaps are closed, there is a
50$\%$ magnitude reduction, namely  $\mathcal{E}=\mathcal{E}_{\text{g}}/2$. When losses are included, the cusp-like feature is rounded and the interaction increases. 

Analytical expressions for the large distance asymptotics  ($d |\tilde{\Delta}|/\hbar c \gg 1$) of the zero temperature Casimir interaction are summarized in Table \ref{Table_GS_SS_Ez_nonzero_A_0} for a given combination of phases in the interacting materials (assumed to have $\mu=0$ and the same $\Gamma$), both for the case of zero and small dissipation.
Each of the entries in the table can be obtained by the following procedure. First, one determines whether a mass gap closes for either of the phases, and then one identifies the corresponding relevant coefficients $B_{-1}^{xx}$, $B_{0}^{xx,xy}$, and $B_{1}^{xx,xy}$. Given this information, one computes the large-distance Casimir energy $\mathcal{E}$ (see Methods) to leading order in the fine structure constant and the distance-decay power, using for the product of the reflection matrices ${\bf R}_1 \cdot {\bf R}_2$ the appropriate combinations of $B$ coefficients for each of the interacting materials. 
Let us first discuss the case of zero dissipation.  When the staggered layers are either in the QSHI or BI phase, all mass gaps are non-zero, the relevant coefficient is $B_{1}^{xx}$ (note that for these phases $B_{0}^{xy}$ vanishes upon summing over valley and spin indices), and the Casimir energy scales as $\mathcal{E}\sim \alpha^2 d^{-5}$. This dependency upon $\alpha^{2}$  suggests a much weaker interaction as compared to two graphene sheets (see Eq. \ref{Eq_Casimir_Graphene}). 
When one of the materials is either in the QSHI or BI phase, while the other one is in the SPM or SVPM phase, two mass gaps are closed, the relevant coefficients are $B_{1}^{xx}$ (for the material in the QSHI/BI phase) and $B_{0}^{xx}=\pi/8$ (for the material in the SPM/SVPM phase), and the energy scales as $\mathcal{E}\sim \alpha^{2} d^{-4}$. 
However, when the SPM or SVPM phase is substituted by an AQHI or PS-QHI phase, the Hall coefficient $B_{0}^{xy}$ becomes relevant, and
the asymptotic Casimir energy is found to be $\mathcal{E}\sim \alpha^{3} d^{-4}$. 
Finite-dissipation corrections $\Delta \mathcal{E}$ to the large-distance Casimir energy are governed by the coefficient  $B_{0}^{xx}$ in all phases. 
Analytical expressions for this correction can be obtained in the limit of small dissipation, 
$\hbar\Gamma \ll \abs{\tilde{\Delta}} \equiv \left(\sum'_{\eta,s=\pm 1} \abs{\Delta^{\eta}_{s}}^{-1} \right)^{-1}$. As shown in Table
\ref{Table_GS_SS_Ez_nonzero_A_0}, $\Delta \mathcal{E}$ inherits the linear in $\Gamma$ dependency from 
$B_{0}^{xx}$ (see Fig. 2(d)), and decays as $d^{-3}$ for all phase combinations. As compared to the lossless case, dissipation results in a qualitative change of the power-law decay of the interaction, in sharp contrast to the situation of typical 3D planar slabs where dissipation only scales the 
large-distance Casimir energy by an overall numerical factor.

%%%%%%%%%%%%%%

\begin{table}[t]
\begin{center}
\renewcommand{\arraystretch}{1.5}
\begin{tabular}{V{3.5} c V{3.5} cV{3.5}cV{3.5}}
\hlineB{3.5}
\rowcolor{cyan!40}[2pt]
 Mat$_{1}$ Mat$_{2}$ & $\mathcal{E}/\mathcal{E}_{\text{g}}(\Gamma = 0)$ & $\Delta\mathcal{E}/\mathcal{E}_{\text{g}}(\hbar\Gamma \ll \abs{\tilde{\Delta}})$ \\ 
\hlineB{3.5}

\scalebox{0.7}{$\begin{array}{c}
\text{QSHI}\\
\text{BI}
\end{array}$}
\scalebox{0.7}{$\begin{array}{c}
\text{QSHI}\\ 
\text{BI}
\end{array}$} & $\frac{(\hbar c)^{2}\alpha}{5\pi\abs{\tilde{\Delta}_{1}}\abs{\tilde{\Delta}_{2}}d^{2}}$ & $\frac{\hbar\Gamma}{3\pi}\frac{\log(\abs{\tilde{\Delta}_{1}}/\abs{\tilde{\Delta}_{2}})}{\abs{\tilde{\Delta}_{1}} - \abs{\tilde{\Delta}_{2}}}$\\
\cline{1-1}

\scalebox{0.7}{$\begin{array}{c}
\text{QSHI}\\
\text{BI}
\end{array}$}
\scalebox{0.7}{$\begin{array}{c}
\text{SPM}\\
\text{SVPM}
\end{array}$} & $\frac{\hbar c\alpha \left[ 1 - 4\log(\pi\alpha/4)\right]}{32\abs{\tilde{\Delta}_{1}}d}$ & $\frac{\hbar\Gamma \log\left (3\pi\abs{\tilde{\Delta}_{1}}/2\hbar\Gamma\right)}{3\pi\abs{\tilde{\Delta}_{1}} - 2\hbar\Gamma}$\\
\cline{1-1}

\scalebox{0.7}{$\begin{array}{c}
\text{QSHI}\\
\text{BI}
\end{array}$}
\scalebox{0.7}{$\begin{array}{c}
\text{AQHI}\\
\text{PS-QHI}
\end{array}$} & $\frac{2 \hbar c \alpha^{2}C_{2}^{2}}{3\pi\abs{\tilde{\Delta}_{1}}d}$ & $\frac{\hbar\Gamma}{3\pi}\frac{\log(\abs{\tilde{\Delta}_{2}}/\abs{\tilde{\Delta}_{1}})}{\abs{\tilde{\Delta}_{2}} - \abs{\tilde{\Delta}_{1}}}$\\
\cline{1-1}

\scalebox{0.7}{$\begin{array}{c}
\text{SPM}\\
\text{SVPM}
\end{array}$}
\scalebox{0.7}{$\begin{array}{c}
\text{SPM}\\
\text{SVPM}
\end{array}$} & $\frac{1}{2}$ & $\frac{\hbar\Gamma}{3\pi\abs{\tilde{\Delta}}}$\\
\cline{1-1}

\scalebox{0.7}{$\begin{array}{c}
\text{AQHI}\\
\text{PS-QHI}
\end{array}$}
\scalebox{0.7}{$\begin{array}{c}
\text{AQHI}\\
\text{PS-QHI}\\
\text{SPM}
\end{array}$} & $\frac{4\alpha}{\pi}C_{1}C_{2}$ & $\frac{\hbar\Gamma}{3\pi}\frac{\log(\abs{\tilde{\Delta}_{1}}/\abs{\tilde{\Delta}_{2}})}{\abs{\tilde{\Delta}_{1}} - \abs{\tilde{\Delta}_{2}}}$\\
\cline{1-1}

\scalebox{0.7}{$\begin{array}{c}
\text{SDC}
\end{array}$}
\scalebox{0.7}{$\begin{array}{c}
\text{AQHI}\\
\text{PS-QHI}
\end{array}$} & $\frac{4\alpha}{\pi}C_{1}C_{2}$ & $\frac{\hbar\Gamma\log\left(3\pi\abs{\tilde{\Delta}_{2}}/4\hbar\Gamma\right)}{3\pi\abs{\tilde{\Delta}_{2}}}$ \\
\cline{1-1}

\scalebox{0.7}{$\begin{array}{c}
\text{SDC}
\end{array}$}
\scalebox{0.7}{$\begin{array}{c}
\text{QSHI}\\
\text{BI}
\end{array}$} &  $\frac{\hbar c\alpha [ 1 - 4\log(\pi\alpha/8) ]}{64\abs{\tilde{\Delta}_{2}}d}$ &  $\frac{\hbar\Gamma\log\left(3\pi\abs{\tilde{\Delta}_{2}}/4\hbar\Gamma\right)}{3\pi\abs{\tilde{\Delta}_{2}}}$ \\
\cline{1-1}

\scalebox{0.7}{$\begin{array}{c}
\text{SDC}
\end{array}$}
\scalebox{0.7}{$\begin{array}{c}
\text{SDC}
\end{array}$} &  $\frac{1}{4}$ &  $\frac{\hbar\Gamma}{3\pi\abs{\tilde{\Delta}}}$ \\

\hlineB{3.5} 
\rowcolor{cyan!40}[2pt]
Mat$_{1}$ Gra & $\mathcal{E}/\mathcal{E}_{\text{g}}(\Gamma=0; \Lambda=0)$   & $\Delta\mathcal{E}/\mathcal{E}_{\text{g}}(\hbar\Gamma \ll \abs{\tilde{\Delta}}; \Lambda=0)$\\ 
\hlineB{3.5} 

\scalebox{0.7}{$\begin{array}{c}
\text{QSHI}\\
\text{BI}
\end{array}$}
\scalebox{0.7}{$\begin{array}{c}
\text{GRA}
\end{array}$} & $\frac{\hbar c \alpha \left[ 1 - 4\log\left(\pi\alpha/2\right)\right]}{16\abs{\tilde{\Delta}_{1}}d}$ & $\frac{\hbar\Gamma\log\left(3\pi\abs{\tilde{\Delta}_{1}}/\hbar\Gamma\right)}{3\pi\abs{\tilde{\Delta}_{1}} - \hbar\Gamma}$\\
\cline{1-1}

\scalebox{0.7}{$\begin{array}{c}
\text{SVPM}
\end{array}$}
\scalebox{0.7}{$\begin{array}{c}
\text{GRA}
\end{array}$} & $\log(2)$ & $\frac{4 \hbar\Gamma}{3\pi\abs{\tilde{\Delta}_{1}} - 2 \hbar\Gamma}\log(2)$\\

\hlineB{3.5}
\end{tabular}
\renewcommand{\arraystretch}{1}
\end{center}
\caption{\label{Table_GS_SS_Ez_nonzero_A_0} 
{\bf Large-distance asymptotics of the zero-temperature Casimir energy in the graphene family.} The left column denotes the phase combinations of the materials (any pair of combinations can be chosen in a given row provided they are realizable for given $\Lambda$ and $E_z$ values), the center column gives the Casimir energy in the lossless case, and the right column provides the correction $\Delta \mathcal{E}$ for small dissipation. When the second layer (Mat$_2$) is graphene, the possible phase combinations for $\Lambda=0$ are shown in the bottom two rows, and when $\Lambda\neq 0$ graphene is in a AQHI phase and the possible phase combinations are given by the 3rd and 5th rows with AQHI for Mat$_2$. The inter-layer separation is large, $d |\tilde{\Delta}|/\hbar c \gg 1$, and
all materials have $\mu=0$ and the same $\Gamma$. 
}
\label{ParamGSGS}
\end{table}
 
%%%%%%%%%%%%%% 
 
A further striking consequence of the different electronic phases in the graphene-family is that the Casimir energy can be quantized. Since the large-distance zero temperature interaction energy between lossless 2D staggered layers in AQHI, PS-QHI, or SPM phases is proportional to their Hall conductivities and $\sigma_{xy}\sim C$, we find that $\mathcal{E}/\mathcal{E}_{\text{g}}= (4 \alpha/\pi) C_1C_2$, i.e. the Casimir energy is quantized in terms of the product of Chern numbers (see Table \ref{Table_GS_SS_Ez_nonzero_A_0}). At this point it is important to emphasize that the reflection matrices entering the Lifshitz formula correspond to reflection of vacuum fluctuations from within the Fabry-P\'{e}rot cavity, and that the sign of the Hall conductivities on either layer (induced by the external circularly polarized laser) changes as seen from fluctuations impinging on the bottom or top layer. The overall result is that the signs of the Chern numbers of the bottom and top layers are different, $C_1 C_2 <0$, and hence the Casimir force is not only quantized but is also repulsive.
This is shown for the case of two dissipationless AQHI or PS-QHI identical sheets in Fig. \ref{fig:Fig3}(e). The zero-temperature Casimir energy  features a ladder-like quantized and repulsive behavior of the Casimir energy $\mathcal{E} \sim - \alpha^2 C_1 C_2 d^{-3} >0$ with the strongest repulsion for $C_1=-C_2=\pm 2$. 
%Note that the repulsive character is independent of whether the incident laser has right or left polarization.
A physical picture of this large-distance Casimir repulsion can be obtained by noting that the polarized laser field induces circulating currents on both layers, whose sense of rotation is determined by the sign of the Hall conductivities. The Casimir cavity is essentially a collection of current loops on each layer facing each other or, equivalently, two parallel sheets of magnetic dipoles. Recalling that  anti-parallel magnetic dipoles repel, it follows that two AQHI, PS-QHI, or SPM layers with Hall conductivities of unequal sign will repel.
For the other SDC/SVPM/SPM phases, on the other hand, the Casimir energy behaves as $\mathcal{E} \sim - \alpha d^{-3} <0$, which
corresponds to the attractive force between two semi-metals, and in the large-distance asymptotics results in an abrupt change of the Casimir force.
The QSHI/BI phases also result in an attractive force but with a stronger decay, $\mathcal{E} \sim - \alpha^2 d^{-5}<0$.
Results for other combination of materials of the graphene family (e.g., silicene-graphene and silicene-germanene) as well as effects of finite dissipation in the quantized repulsive Casimir force are shown in the Supplementary Note 3.

We briefly discuss the effect of the chemical potential. As long as $\abs{\mu} < \abs{\Delta_{s}^{\eta}}$ for all Dirac cones, the results described above for the $\mu=0$ case still hold. When $\abs{\mu} > \abs{\Delta_{s}^{\eta}}$ for at least one Dirac cone, 
the intraband conductivity (Eq.\ref{sigma_Dirac_Cone}) starts to play a role. For $\Gamma=0$, the low-frequency optical response is dominated by the plasma-like term in Eq.\ref{ConductivityExpansion} containing $B_{-1}^{xx}$, and the large-distance Casimir energy corresponds to that of a perfect conductor. For $\Gamma>0$,  $B_{-1}^{xx}=0$ and the dominant contribution to the large-distance Casimir energy comes from $B_{0}^{xx}$. 
In the limit $|\mu| \gg \hbar \Gamma$, the Casimir energy corresponds to that of 2D Drude metals. 
Further details of the effect of $\mu$ on Casimir force phase transitions can be found in the Supplementary Note 3.\\

\hspace{-12pt}
{\bf Thermal corrections to the Casimir energy.} Thermal effects enter in the Lifshitz formula by replacing the integral over complex frequencies with a summation over Matsubara frequencies and taking into account the finite-temperature conductivity (see Methods and Supplementary Note 4).
In Fig. 4(a) we show the Casimir energy between identical layers of the graphene family as a function of temperature at a fixed distance  for some representative points in phase space.  
For low temperatures $k_{\text{B}} T / \lambda_{\text{SO}} \ll 10^{-3}$, Casimir repulsion for the CI-CI phases is still present (dashed blue curve), and as the temperature increases there is a cross-over to attraction. Another effect of temperature is to reduce the contrast in magnitude between Casimir energies for different points in phase space (e.g. SPM and QSHI phases shown in green and black), which ultimately results in blurred Casimir force phase transitions. For large temperatures 
$k_{\text{B}}T / \lambda_{\text{SO}} \gtrsim 10^{-2}$, all curves are essentially described by the classical limit $\mathcal{E}_{n=0}=- \zeta(3)k_{\text{B}}T/16 \pi d^2$, 
which is the same for all points in phase space (see Supplementary Note 4). Fig. 4(b) depicts the distance dependence of the Casimir energy for the QSHI phase for various temperatures, showing a change of scaling law from $\mathcal{E} \sim \alpha d^{-3}$ for 
$d \lambda_{\text{SO}}/\hbar c  \ll 10^{-2}$ to $\mathcal{E} \sim T d^{-2}$ for 
$d \lambda_{\text{SO}}/\hbar c  \gg 1$. Figs. 4(c) and (d) show different cuts of the Casimir energy phase diagram for fixed temperature and distance.
Thermal effects result in the smoothing out of phase transition boundaries and disappearance of quantized and repulsive Casimir interactions. 
For example, for the case of a stanene cavity maintained at  liquid helium temperature $T=4.2$ K (corresponding to
$k_{\text{B}}T / \lambda^{\text{Stan}}_{\text{SO}} =3.6 \times 10^{-3}$), Casimir force phase transitions are still observable in the smoothed cusp-like features. \\

\begin{figure}[t]
\centering
\includegraphics[width=1\linewidth]{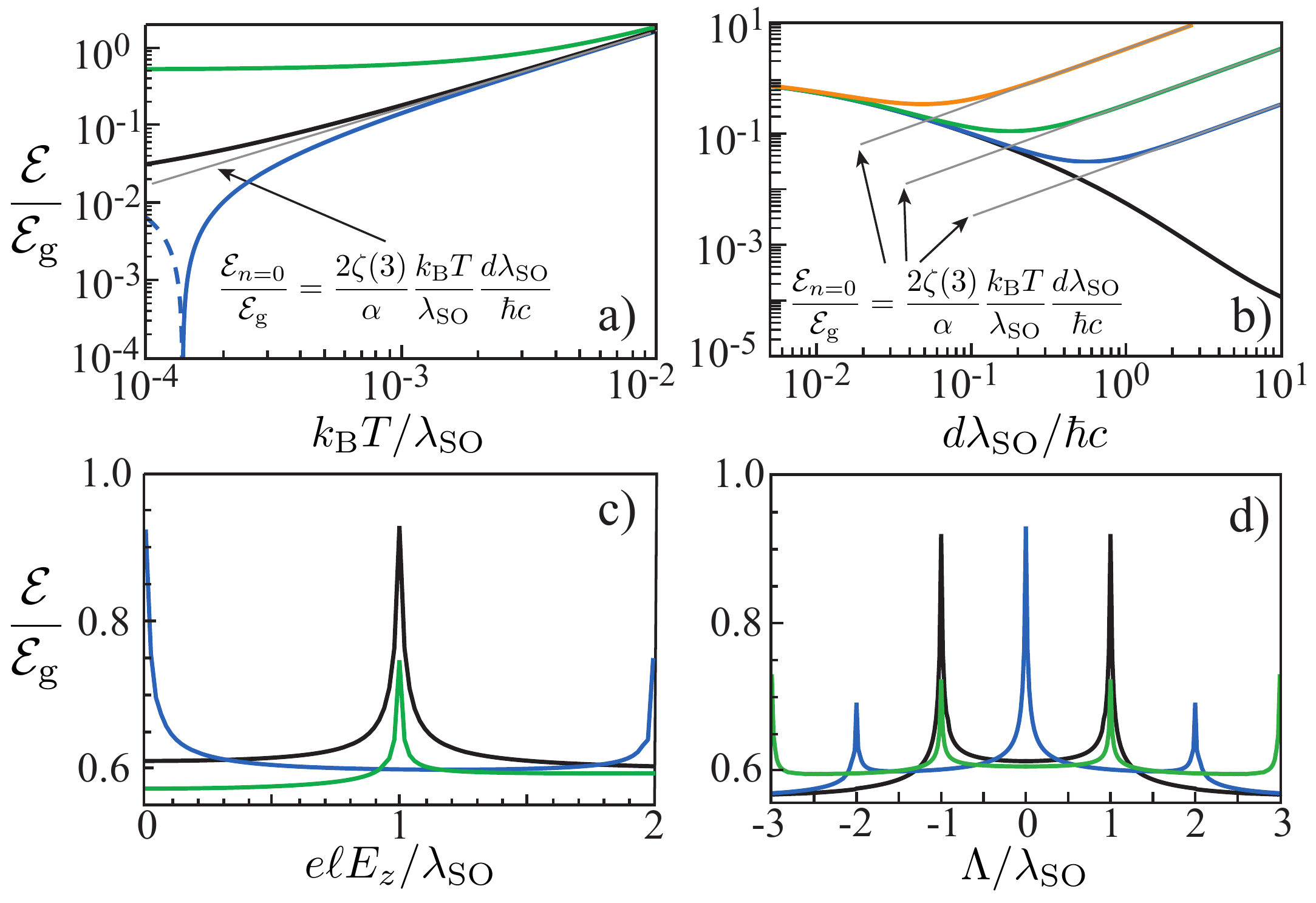}
\caption{
{\bf Finite-temperature Casimir interaction in the graphene family.}
(a) Casimir energy between identical layers of the graphene family as a function of temperature for some representative points 
$(e \ell E_z/\lambda_{\text{SO}},\Lambda/\lambda_{\text{SO}})$
in phase space: $(0,0)$ (black), $(0,1)$ (green), and $(0,2)$ (blue;
dashed blue corresponds to repulsion, where $-\mathcal{E}/\mathcal{E}_{\text{g}}>0$).
The distance is fixed at $d \lambda_{\text{SO}}/\hbar c = 0.5$. 
(b) Distance dependency of the Casimir energy at the origin of phase space 
$E_z=\Lambda=0$ for various temperatures $k_{\text{B}}T /\lambda_{\text{SO}}=0$ (black), $10^{-4}$ (blue), $10^{-3}$ (green), and $10^{-2}$ (orange).
The thing gray curves in (a) and (b) denote the $n=0$ Matsubara contribution to the energy (see Supplementary Information).
(c) Cut of the Casimir energy phase diagram along $\Lambda/\lambda_{\text{SO}}=0$ (black), $1$ (blue), and $2$ (green) as a function of the electric field. (d) Cut along $e \ell E_z/\lambda_{\text{SO}}=0$ (black), $1$ (blue), and $2$ (green) as a function of $\Lambda/\lambda_{\text{SO}}$. In both (c) and (d) the temperature is 
$k_{\text{B}}T / \lambda_{\text{SO}} = 3.6 \times 10^{-3}$ (corresponding to $4.2$ K for stanene), and the distance is fixed at
$d \lambda_{\text{SO}}/\hbar c = 0.5$ (corresponding to $940$ nm for stanene).
In all plots $\mu=0$, $\hbar \Gamma/\lambda_{\text{SO}} = 10^{-4}$, and the normalization $\mathcal{E}_{\text{g}}$ is the zero-temperature graphene energy given in Eq.(\ref{Eq_Casimir_Graphene}).
}
\label{fig:Fig4}
\end{figure}

%%%%%%%%%%%%%%

\hspace{-12pt}
{\bf Discussion}

\hspace{-12pt} Our study shows that in order to probe the Dirac-like physics in the graphene family via the rich structure of its Casimir interactions, low temperature set-ups, such as the cryogenic AFM developed in \cite{Laurent2012} to measure Casimir force gradients using a metallic spherical tip, are required. In order to suggest possible experimental signatures of the Casimir force phase transitions, let us consider a stanene layer (neutral and weakly dissipative $\hbar \Gamma/\lambda^{\text{Stan}}_{\text{SO}}=10^{-4}$) under varying static field along the $\Lambda=0$ line in Fig. 1(b) in front of bulk gold semi-infinite substrate. Evaluating the Casimir pressure at liquid helium temperature $T=4.2$ K and at a distance of 
$d=100$ nm, we obtain $P_{\text{QSHI-Au}} \simeq 0.2$ Pa at $E_z=0$ and  $P_{\text{SVPM-Au}} \simeq 0.3$ Pa
at $e \ell E_z/\lambda^{\text{Stan}}_{\text{SO}}=1$. In the proximity force approximation (valid for $d \ll R$, where $R$ is the radius of curvature of the metallic sphere), the respective Casimir force gradients $F'/R \approx 2 \pi P$ are $1.3$ and $1.9$ Pa. Given the reported sensitivities for $F'/R$ of $0.1$ Pa \cite{Laurent2012}, it should be possible to probe Casimir force phase transitions in this set-up.

It is worth noting that when there is an applied polarized laser ($\Lambda \neq 0$), there is an additional optical force on top of the Casimir interaction between the two parallel layers of the graphene family.
A straightforward calculation of the optical pressure to leading order in $\alpha$ gives
%$P_{opt}\simeq (I_0 \pi \alpha/c) [1-\pi \alpha/2 + (\pi \alpha)^4 \cos^2(\omega_0 d/c)/4]$
$P_{\text{opt}}\simeq I_0 \pi \alpha/c$, where $I_0$ is the laser intensity.
Typical laser parameters for which the low-energy Hamiltonian Eq. \ref{Low_Energy_Hamiltonian}
is valid and for which the phase diagram in Fig. 1(b) can be explored, result in an optical force larger than the Casimir one.  Nevertheless, it is still possible to distinguish between the two forces by taking advantage of the particular dependency of the optical force on the laser parameters. For example, modulating the laser polarization between circular ($\Lambda \neq 0$) and linear ($\Lambda= 0$, since linearly polarized light does not break time reversal symmetry \cite{Kane-2009}) states, 
and employing a lock-in technique at the modulation frequency, the optical force is removed from the signal (as it is independent of the state of polarization), and one can detect the difference between the Casimir force at $(E_z,\Lambda)$ and at $(E_z,0)$. This measurement, in conjunction with an independent detection of the force for no applied laser field, allows the determination of the Casimir force as a function of distance at any point $(E_z,\Lambda)$ in the phase diagram, irrespective of the strength of the optical force. 

We have shown that the Casimir interaction in materials of the graphene family has a rich structure due to their unique electronic and optical properties. Their various electronic phases, tunable by external fields, result in Casimir force phase transitions featuring different distance scaling laws, significant magnitude changes, force quantization and repulsion. The measurement of some of these effects should be within reach with current state-of-the-art low-temperature Casimir force experiments.\\

%%%%%%

\hspace{-12pt}
{\bf Methods}

\hspace{-12pt}
The Casimir interaction energy per unit area between two parallel plates separated by a distance $d$ at temperature $T$ can be calculated using the Lifshitz formula \cite{Woods-RevModPhys,RPM_Galina_Klimchitskaya}
\[
\mathcal{E}(T)= k_{\text{B}}T {\sum_{n}}' \int\frac{d^{2}{\bf k}_{\|}}{(2\pi)^2} \log \det \left( {1 - {\bf R}_{1}\cdot {\bf R}_{2} e^{-2 k_{z,n} d}}\right), 
\]
where the summation is over Matsubara frequencies $\xi_n=2 \pi n k_{\text{B}}T /\hbar$ ($n=0,1,2,\ldots$), and the prime indicates that the $n=0$ term has a $1/2$ weight.  Furthermore, $k_{z,n} = \sqrt{\textbf{k}^{2}_{\|} + \xi_n^{2}/c^{2}}$ and ${\bf R}_{1,2} = {\bf R}_{1,2} (i \xi_n, {\bf k}_{\|})$ are $2\times2$ reflection matrices. The $T=0$ formula is obtained by the replacement
$k_{\text{B}}T {\sum_{n}}' \rightarrow (\hbar/2 \pi) \int_0^{\infty} d\xi$.
The diagonal elements of the reflection matrices are the $R_{\text{ss}}$ and $R_{\text{pp}}$ Fresnel coefficients, and the off-diagonal 
elements $R_{\text{sp}, \text{ps}}$ arise from the Hall conductivity that induces polarization conversion. 
Imposing standard boundary conditions to Maxwell's equations for a single 2D sheet, one finds \citep{CasimirCI}
\begin{eqnarray}
R_{\text{ss}} & = & - \frac{2\pi}{\delta_n}\left[ \frac{\sigma_{xx}}{c\lambda_n} + \frac{2\pi}{c^2}\left( \sigma_{xx}^{2} + \sigma_{xy}^{2} \right)\right], \nonumber  \\
R_{\text{sp}} & = &   R_{\text{ps}} = \frac{2\pi \sigma_{xy}}{\delta_n c}, \nonumber \\
R_{\text{pp}} & = &   \frac{2\pi}{\delta_n}\left[ \lambda_n\dfrac{\sigma_{xx}}{c} + \frac{2\pi}{c^2}\left( \sigma_{xx}^{2} + \sigma_{xy}^{2} \right)\right], 
\end{eqnarray}
where $\delta_n=1 + 2\pi\frac{\sigma_{xx}(1+\lambda_n^2)}{c\lambda_n} + \frac{4\pi^{2}}{c^2}\left( \sigma_{xx}^{2} + \sigma_{xy}^{2} \right)$, 
$\lambda_n = k_{z,n}c/\xi_n$, and the conductivity tensor is evaluated at the imaginary Matsubara frequencies $\sigma_{ij}(i \xi_n)$.
Note that in the Lifshitz formula the Hall conductivities on either plate must have opposite signs since ${\bf R}_{j}$ correspond to reflections within the 
Fabry-P\'{e}rot cavity.  
In the estimation of the Casimir pressure between stanene and a gold bulk, we model the permittivity of Au as $\epsilon_{\text{Au}}(i\xi) = 1 + \Omega_{\text{p}}^{2}/(\xi^{2} + \xi \gamma_{\text{p}}) + \chi_{0}\xi_{0}^{2}/(\xi^{2} + \xi^{2}_{0} + \xi \gamma_{0})$, with $(\Omega_{\text{p}},\gamma_{\text{p}},\xi_{0},\gamma_{0})=(13.7, 0.05, 20, 25) \times 10^{15} \text{rad } \text{s}^{-1}$ and $\chi_{0}=5$ \cite{Laurent2012}. \\

\hspace{-12pt}
{\bf Data availability.} The data that support these findings are available from the corresponding authors on request.\\

\hspace{-12pt}
{\bf Acknowledgments}

\hspace{-12pt}
We acknowledge financial support from the US Department of Energy under grant No. DE-FG02-06ER46297,  LANL LDRD program, and CNLS. P.R.-L. also acknowledges partial support from TerMic (Grant No. FIS2014-52486-R, Spanish Government). We are grateful to Ricardo Decca for insightful discussions.\\

\hspace{-12pt}
{\bf Author contributions.} 

\hspace{-12pt} All authors contributed equally to this work.\\

\hspace{-12pt}
{\bf Additional information} 

\hspace{-12pt} {\bf Supplementary information} is available in the online version of the paper.\\

\hspace{-12pt}
{\bf Competing financial interests:} The authors declare no conflict of interest.\\

\textsuperscript{1} Correspondence: lmwoods@usf.edu, dalvit@lanl.gov\\

%%%%%%%%%%%%%%%%%%%%%%

\pagebreak

\newcounter{defcounterfigure}
\setcounter{defcounterfigure}{0}
\newenvironment{myfig}{%
\addtocounter{figure}{0}
%\addto\captionsenglish{\renewcommand{\figurename}{Supplementary Figure}}
\refstepcounter{defcounterfigure}
\renewcommand\thefigure{\thedefcounterfigure}
\begin{figure}}
{\end{figure}}

\begin{widetext}

\hspace{-12pt}
{\bf Supplementary Note 1. Zero temperature optical conductivity.}

\hspace{-12pt} The components of the optical conductivity tensor of each Dirac cone $\sigma_{ij}(i \xi, \Delta^{\eta}_s,\mu,T=0)$
can be written as an asymptotic series in the limit of small frequencies, as shown in Eqs. (4) in the main text. 
The corresponding coefficients $B_{-1}^{xx}$, $B_{0}^{xx,xy}$, and $B_{1}^{xx,xy}$ have intra- and inter-band contributions, i.e.,
$B_i^{xx} = B_i^{xx, {\rm intra}} + B_i^{xx,{\rm inter}}$ and $B_i^{xy} = B_i^{xy, {\rm intra}} + B_i^{xy,{\rm inter}}$ (here, $i = 0, \pm 1$).  
The intra-band coefficients are given by
\begin{myeqnarray}
B_{-1}^{xx,\text{intra}} &=& \dfrac{\mu^{2} - {(\Delta_{s}^{\eta}})^{2}}{|\mu|}\dfrac{\delta_{\Gamma,0}}{2\hbar}\Theta(|\mu| - |\Delta_{s}^{\eta}|) \, , \nonumber \\
B_{0}^{xx,\text{intra}} &=& \dfrac{\mu^{2} - ({\Delta_{s}^{\eta}})^{2}}{|\mu|}\dfrac{1 - \delta_{\Gamma,0}}{2\hbar\Gamma}\Theta(|\mu| - |\Delta_{s}^{\eta}|)\, , \nonumber \\
B_{1}^{xx,\text{intra}} &=& \dfrac{\mu^{2} - ({\Delta_{s}^{\eta}})^{2}}{|\mu|}\dfrac{1 - \delta_{\Gamma,0}}{2\hbar\Gamma^{2}}\Theta(|\mu| - |\Delta_{s}^{\eta}|)\, ,\nonumber \\ 
B_{0}^{xy,\text{intra}}& = &  B_{1}^{xy,\text{intra}} =0 ,
\label{sup1}
\end{myeqnarray}
where $\delta_{\Gamma,0}$ is the Kronecker delta and $\Theta(\Gamma)$ is the Heaviside step function. The inter-band coefficients are
\begin{myeqnarray}
B_{-1}^{xx,\text{inter}} & =& 0, \nonumber \\
B_{0}^{xx,\text{inter}} & = &\dfrac{({\Delta_{s}^{\eta}})^{2}}{2M\hbar\Gamma} + \left( \dfrac{1}{4} - \dfrac{({\Delta_{s}^{\eta}})^{2}}{\hbar^{2}\Gamma^{2}} \right) \tan^{-1}\left(\dfrac{\hbar\Gamma}{2M}\right),  \nonumber \\
B_{1}^{xx,\text{inter}}& = & \dfrac{\hbar^{2}\Gamma^{2}M^{2} - \hbar^{2}\Gamma^{2}({\Delta_{s}^{\eta}})^{2} - 8({\Delta_{s}^{\eta}})^{2}M^{2}}{8\hbar\Gamma^{2}M^{3} + 2\hbar^{3}\Gamma^{4}M} + \dfrac{2({\Delta_{s}^{\eta}})^{2}}{\hbar^{2}\Gamma^{3}}\tan^{-1}\left(\dfrac{\hbar\Gamma}{2M}\right),  \nonumber \\
B_{0}^{xy,\text{inter}}& = & \dfrac{\eta\Delta_{s}^{\eta}}{\hbar\Gamma}\tan^{-1}\left(\dfrac{\hbar\Gamma}{2M}\right),  \nonumber \\
B_{1}^{xy,\text{inter}}& = & \dfrac{\eta\Delta_{s}^{\eta}}{\hbar\Gamma^{2}}\left[\dfrac{2 \hbar\Gamma M}{\hbar^{2}\Gamma^{2} + 4M^{2} } - \tan^{-1}\left(\dfrac{\hbar\Gamma}{2 M}\right)\right]\, .
\label{sup2}
\end{myeqnarray}
These expressions show how the individual contributions in the longitudinal and Hall conductivities of the graphene family materials depend on the Dirac mass of a particular cone, the chemical potential and scattering rate. They are especially useful for understanding how the internal properties and external factors, such as the applied electric field and circularly polarized light, determine the low-frequency electro-optical response of these materials. In Fig. 2(d) of the main text we plot these coefficients as a function of $\hbar \Gamma/|\Delta^{\eta}_s|$ for the case of neutral layers ($\mu=0$). The full conductivity tensor is obtained by summing over spin and valley indices, and can be also written as in Eq. 4 of the main text with coefficients $\tilde{B}_i^{xx,xy} = \sum_{\eta,s} B_i^{xx,xy}$.

Equations \ref{sup1} and \ref{sup2} contain all the necessary information needed to determine the large-distance asymptotics of the zero-temperature Casimir energy between two layers of 2D staggered materials, shown in Table 1 of the main text. Using these low-frequency expansions of the conductivity tensor in the expressions for the reflection coefficients, and computing the Casimir energy as given by the Lifshitz formula to leading order in the fine structure constant, we can obtain the various entries of the table. Different phase combinations of the two layers forming the 
Fabry-P\'{e}rot cavity determine which $B$ coefficients give the dominant contribution to the large-distance scaling law for the Casimir energy, as explained in the main text.
\\

%%%%%%%%

\hspace{-12pt}
{\bf Supplementary Note 2. Finite temperature optical conductivity.}

\hspace{-12pt}
At finite temperatures the optical conductivity can be calculated via the Maldague formula \cite{Giuliani,Maldague}
\begin{myeqnarray}
\label{sigma_T}
\sigma_{ij}(i\xi,\Delta^{\eta}_s,\mu, T) = \int_{-\infty}^{\infty}dE\dfrac{\sigma_{ij}(i\xi,\Delta^{\eta}_s, E, 0)}{4 k_{\rm B}T\cosh^{2}\left(\frac{E-\mu}{2 k_{\rm B}T}\right)}\, ,
\end{myeqnarray}
where $\sigma_{ij}(i\xi,\Delta^{\eta}_s, E,0)$ is the zero-temperature conductivity studied in the previous Section. In Supplementary Figure \ref{Cond1} we show the longitudinal and Hall conductivites $\sigma_{ij}(i\xi,\mu,T)=\sum_{\eta,s} \sigma_{ij}(i\xi,\Delta^{\eta}_s, \mu,T)$
along imaginary frequencies 
for various temperatures. Panels (a-c) are the finite-temperature versions of the zero-temperature panels in Figs. 2(a-c) in the main text.
For low temperatures $k_{\rm B} T/ \lambda_{\rm SO} \ll 1$, the conductivity is essentially identical to the one at zero temperature (see Supplementary Figures \ref{Cond1}(a,b)), except for regions in the phase-space diagram where gaps close (Supplementary Figure \ref{Cond1}(c)). As temperature is increased, thermal effects become relevant at low frequencies, and they mainly affect the longitudinal conductivity.

\begin{myfig}[!h]
\centering 
\includegraphics[width=1\linewidth]{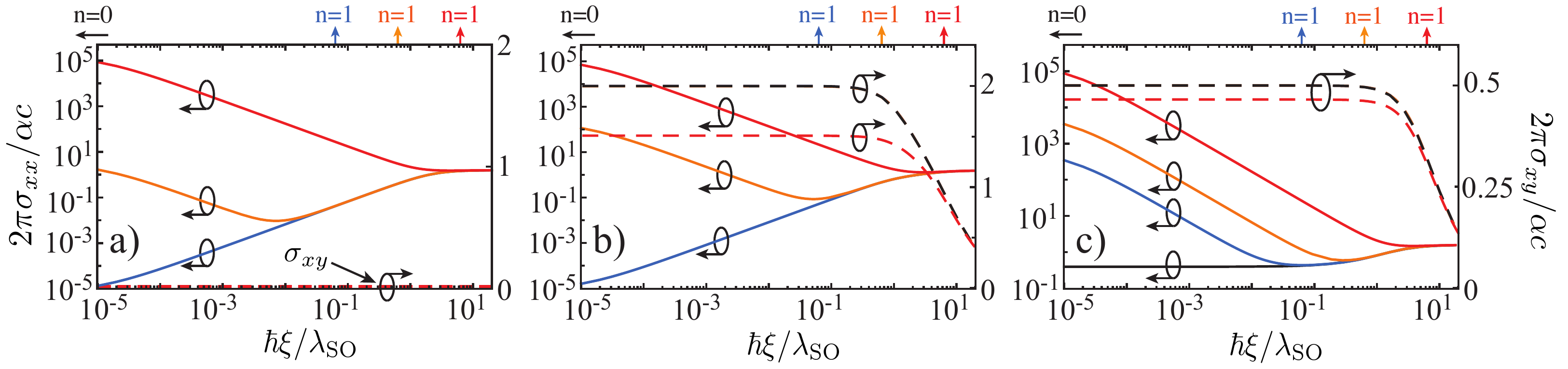}
\caption{
{\bf Temperature dependency of the longitudinal (solid lines) and Hall (dashed lines) conductivities at imaginary frequencies.} Temperatures $k_{\rm B} T/\lambda_{\rm SO}$ are $0$ (black), $10^{-2}$ (blue), $10^{-1}$ (orange), and $1$ (red). The behaviour of the conductivities along imaginary frequencies for different phases is shown: (a)  $E_z=\Lambda=0$ (QSHI phase with $C=0$); (b) $\Lambda/\lambda_{\rm SO}=-3/2$ and $E_{z}=0$ (AQHI phase with $C=2$); and (c) $e \ell E_z/\lambda_{\rm SO}=-\Lambda/\lambda_{\rm SO}=1/2$ (SDC phase with $C=1/2$). Due to the chosen scale, some curves are on top of each other (solid and dashed black and blue in (a) and (b); dashed black, blue, and orange in (b) and (c)).
The values of the longitudinal and Hall conductivities at the $n=0$ Matsubara are approximately equal to those corresponding to the smallest frequency shown. The position of the $n=1$ Matsubara frequency for each temperature is shown on the top of each panel. In all cases $\mu=0$ and $\hbar \Gamma/\lambda_{\rm SO} = 10^{-5}$. 
}
\label{Cond1}
\end{myfig} 
\vspace{2.5cm}

In connection with the computation of the finite-temperature Casimir energy, one can see that the main effect of temperature is on the zero Matsubara frequency $\xi_{n=0}$. The conductivity at all other Matsubara frequencies $n \ge 1$ is basically unaffected by temperature. In Supplementary Figure \ref{Cond2} we show the variation with temperature of $\sigma_{ij}$ at the zero Matsubara in all phase-space. Note that the main effect of temperature is to blur the phase transition boundaries, especially on the longitudinal conductivity. \\

\begin{myfig}[!h]
\centering 
\includegraphics[width=1\linewidth]{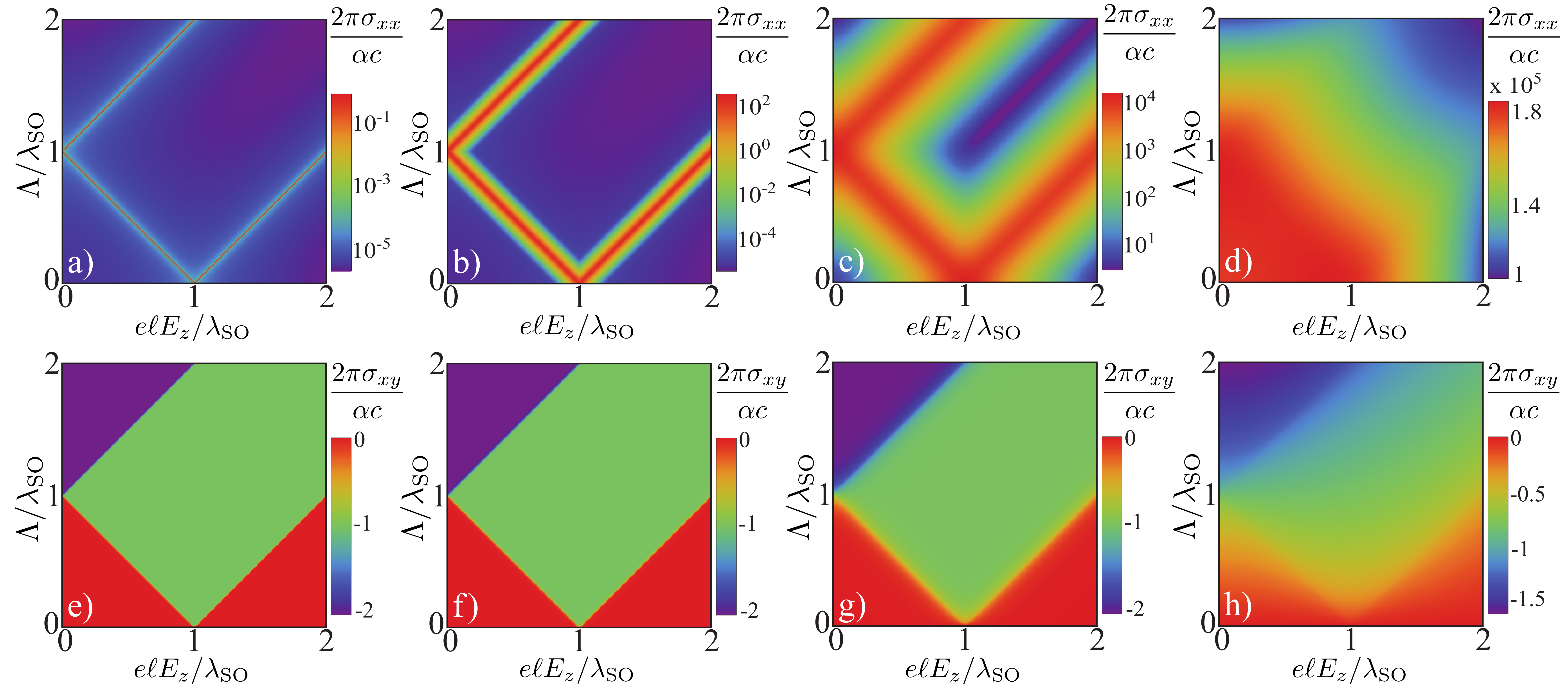}
\caption{
{\bf Phase diagram for the longitudinal (top panels) and Hall (bottom panels) conductivities at zero Matsubara frequency}. Temperature is  $k_{\rm B} T/\lambda_{\rm SO}$: $0$ (a,e), $10^{-2}$ (b,f), $10^{-1}$ (c,g), and $1$ (d,h). In all cases $\mu=0$ and $\hbar \Gamma/\lambda_{\rm SO} = 10^{-5}$. 
}
\label{Cond2}
\end{myfig}

\hspace{-12pt}
{\bf Supplementary Note 3. Zero-temperature Casimir energy in the graphene family.}

\hspace{-12pt}
When the chemical potential satisfies  $\abs{\mu} < \abs{\Delta_{s}^{\eta}}$ for all Dirac cones, all intra-band $B$ coefficients are identically zero, and the Casimir interaction is the same as that of the neutral case $\mu=0$. When $\abs{\mu} > \abs{\Delta_{s}^{\eta}}$ for at least one Dirac cone, the intra-band coefficients become important. Specifically, when $\Gamma=0$ we get $B_{-1}^{xx,{\rm intra}} \neq 0$, and when $\Gamma\neq 0$ we get $B_{-1}^{xx,{\rm intra}} = 0$ and $B_{0}^{xx,{\rm intra}} > 0$. These properties are then reflected in the full coefficients $\tilde{B}_{i}^{xx}$ after summation over spin and valley indices.
When $\tilde{B}_{-1}^{xx} > 0$, this coefficient dominates the low-frequency reflection properties of the layer, which shows a plasma-like metallic behavior. When dissipation is taken into account, the $\tilde{B}_{0}^{xx}$ coefficient is the first relevant one, and the layer has a Drude-like metallic behavior as long as $|\mu| \gg \hbar \Gamma$.

Supplementary Figure \ref{ChemicalPotential} shows the zero-temperature Casimir energy phase diagram for two identical layers of the graphene family separated by $d\lambda_{\rm SO}/\hbar c=1$ for $\hbar\Gamma/\lambda_{\rm SO} = 0.0025$. Each of the panels depicts the evolution of the Casimir energy (normalized by the one for two neutral graphene sheets $\mathcal{E}_{\rm g} = -\hbar c \alpha/32 \pi d^3$)  as the chemical potential increases. Panel \ref{ChemicalPotential}(a) corresponds to neutral layers, $\mu = 0$, and is qualitatively the same as Supplementary Figure 3(b) of the main text. All other panels Supplementary Figure  \ref{ChemicalPotential}(b-d) satisfy $\hbar \Gamma \ll |\mu|$ and, therefore, correspond to the small dissipation limit.  
For $\mu/\lambda_{\rm SO} < 1$ (\ref{ChemicalPotential}(b)) the phase diagram resembles the one of neutral layers as long as all the mass gaps are larger than the Fermi energy (for instance, close to $E_z = \Lambda = 0$). 
In contrast, in regions where $|\Delta^{\eta}_s|<|\mu|$ the Casimir energy is largely increased due to the intra-band conductivity (orange bands in the phase diagram). For $\mu/\lambda_{\rm SO} = 1$ all points in the shown phase diagram \ref{ChemicalPotential}(c) have at least one mass gap smaller than the chemical potential,  $\abs{\Delta_{s}^{\eta}}< \abs{\mu} $, except along the line $\Lambda/\lambda_{\rm SO} = \ell E_z/\lambda_{\rm SO} \geq 1$ where  $\abs{\Delta_{s}^{1}} = \abs{\mu} $ and $\abs{\Delta_{s}^{-1}} > \abs{\mu}$. Finally, for $\mu/\lambda_{\rm SO} = 2$ (panel \ref{ChemicalPotential}(d)) the chemical potential is larger than all mass gaps in the whole phase diagram shown in the figure.  

\begin{myfig}[!h]
\centering 
\includegraphics[width=1\linewidth]{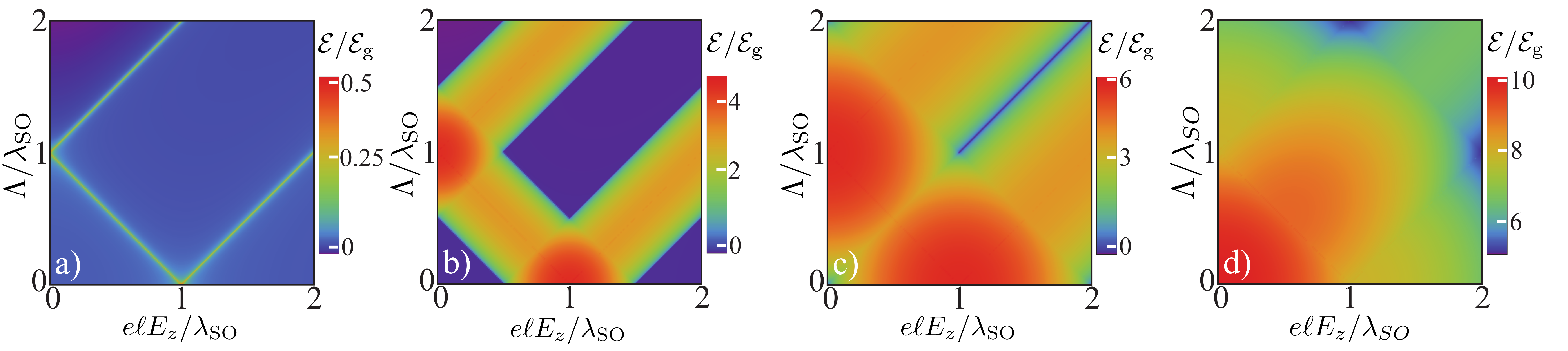}
\caption{
{\bf Normalized zero-temperature Casimir energy $\mathcal{E}/\mathcal{E}_{\rm g}$ phase diagram for two parallel layers for various chemical potentials.} $\mu/\lambda_{\rm SO}$ is equal to $0$ (a), $0.5$ (b), $1$ (c), and $2$ (d). Parameters are  $\hbar\Gamma/\lambda_{\rm SO} = 0.0025$ and $d\lambda_{\rm SO}/\hbar c = 1$.
}
\label{ChemicalPotential}
\end{myfig} 

Supplementary Figure \ref{CasimirLosses} shows how the  Casimir energy phase diagram of Supplementary Figure 3(b) in the main text is modified due to finite dissipation in the materials.
Losses result in the blurring of the phase boundaries and, for sufficiently large dissipation, the Casimir force is attractive for all phases.
\begin{myfig}[!h]
\centering 
\includegraphics[width=0.3\linewidth]{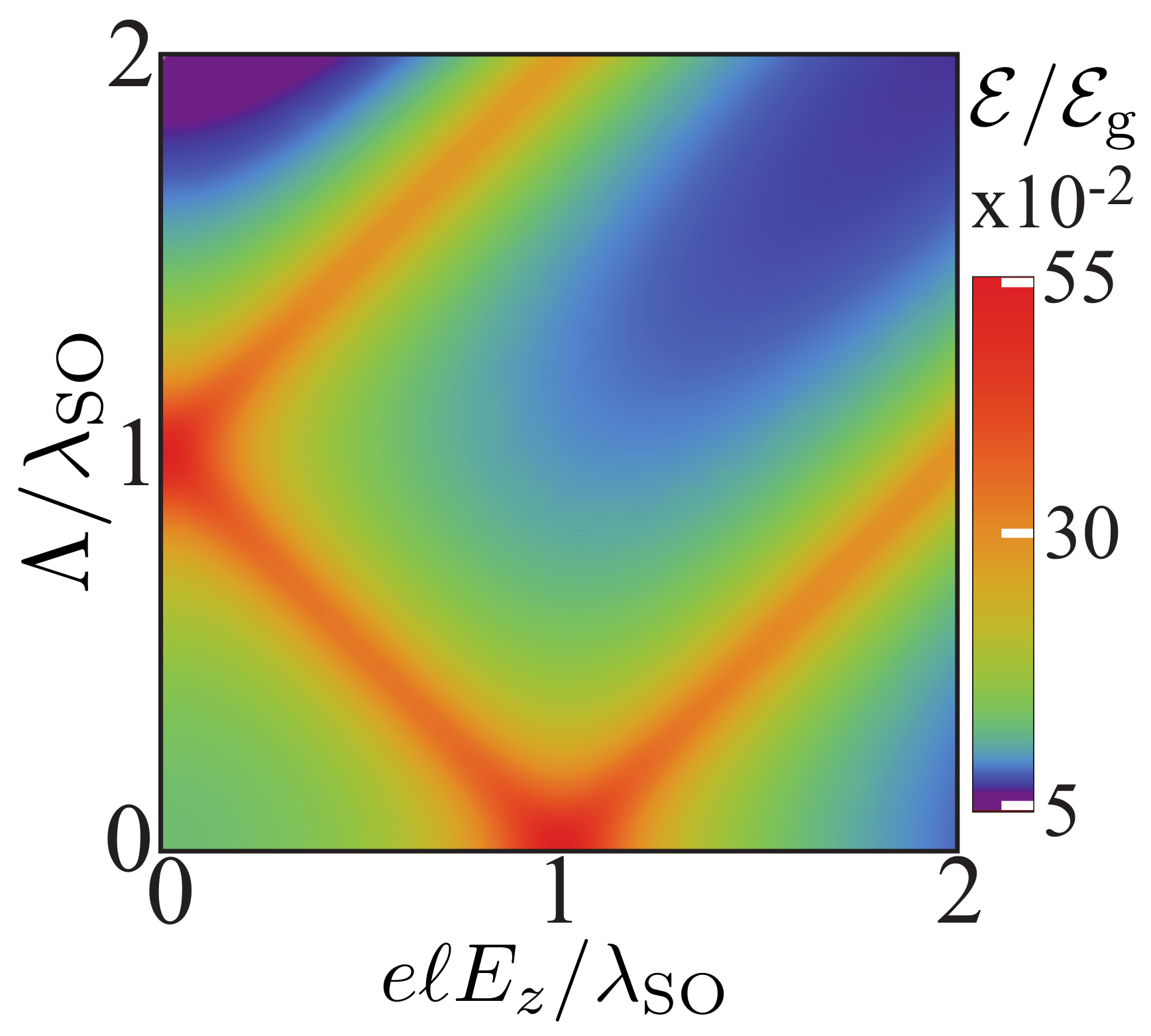}
\caption{
{\bf Zero-temperature Casimir energy phase diagram for two identical parallel layers of the graphene family with finite dissipation}. Parameters are  
$\hbar \Gamma/\lambda_{\rm SO}=0.25$, $d=\hbar c/\lambda_{\rm SO}$, and $\mu=0$.
}
\label{CasimirLosses}
\end{myfig} 

We discussed in the main text the possibility of Casimir force quantization and repulsion between identical layers of the graphene family. Analogous effects take place for two dissimilar layers, as shown in Supplementary Figure \ref{RepulsionLosses}(a)
for the case of dissipationless AQHI/PS-QHI/SPM silicene and AQHI graphene phase combinations, and 
in Supplementary Figure \ref{RepulsionLosses}(b) for silicene-germanene in AQHI, PS-QHI, or SPM phase combinations. Both 
feature a ladder-like quantized and repulsive behavior of the Casimir energy $\mathcal{E} \sim - \alpha^2 C_1 C_2 d^{-3} >0$ with the strongest repulsion for $C_1=-C_2=\pm 2$.  All other phase combinations  result in a stronger decay with distance ($\sim d^{-4},  d^{-5}$), except for the case of 
silicene-graphene close to $\Lambda=0$ and $\ell E_{z}/\lambda^{\rm Sil}_{\rm SO} =1$, for which $\mathcal{E} \sim - \alpha d^{-3} <0$.
This corresponds to the attractive force between two semi-metals (SVPM silicene-graphene), and in the large-distance asymptotics results in an abrupt change of the Casimir force.
Results for finite $\Gamma$ are also shown in Supplementary Figures \ref{RepulsionLosses}(c,d). Small dissipation leads to less well defined boundaries between the different phases with small peaks appearing at the steps of the ladder. Further increasing $\Gamma$ makes the interaction attractive. As the interacting layers are brought closer together, the longitudinal conductivities become important ultimately resulting in Casimir attraction at shorter separations.\\

\begin{myfig}[!h]
\centering 
\includegraphics[width=1\linewidth]{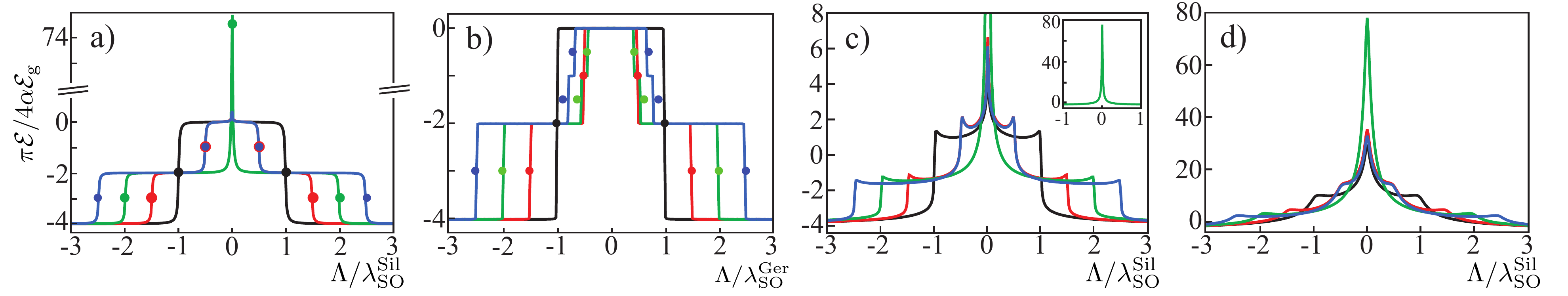}
\caption{
{\bf Quantized Casimir energy at zero temperature.} The normalized Casimir energy $\pi \mathcal{E}/4 \alpha \mathcal{E}_{\rm g}$ is shown as a function of $\Lambda$ 
and $\ell E_{z}/\lambda_{\rm SO} = \{0, 1/2, 1, 3/2\}$ (black, red, green and blue curves, respectively)
for neutral and dissipationless (a) silicene-graphene ($d \lambda_{\rm SO}^{\rm Sil}/\hbar c=10$) and (b) 
silicene-germanene ($d \lambda_{\rm SO}^{\rm Ger}/\hbar c=10$). In the large-distance asymptotics (as given in Table I of the main text), the rounded plateaus become abrupt jumps and the interaction energies 
at phase transition boundaries are the dots in-between plateaus.
The influence of dissipation for the silicene-graphene case is shown in (c) for $\hbar \Gamma/\lambda_{\rm SO}^{\rm Sil} =0.025$ 
and (d) $\hbar \Gamma/\lambda_{\rm SO}^{\rm Sil} =0.25$. The inset in (c) is a zoom-in of the energy around $\Lambda=0$.}
\label{RepulsionLosses}
\end{myfig} 
%\vspace{1.5cm}

\hspace{-12pt}
{\bf Supplementary Note 4. Finite-temperature Casimir energy in the graphene family.}

\hspace{-12pt}
At finite temperature, the Casimir interaction energy between two layers of the graphene family is given by
\begin{myeqnarray}
\mathcal{E}(T)= \frac{k_{\rm B} T}{2 \pi d^2}  {\sum_{n}}' \int_0^{\infty} d\tilde{k} \; \tilde{k}  
\log \det 
\left[ 1 - {\bf R}_{1}(c \tilde{\xi}_n/d, \tilde{k}/d) \cdot {\bf R}_{2}(c \tilde{\xi}_n/d, \tilde{k}/d)
\; e^{-2 \sqrt{\tilde{k}^2+ \tilde{\xi}^2_n}} \right],
\end{myeqnarray}
where the summation is over dimensionless Matsubara frequencies $\tilde{\xi}_n=2 \pi n k_{\rm B} T d /\hbar c$ ($n=0,1,2,\ldots$), the prime indicates that the $n=0$ term has a $1/2$ weight, and $\tilde{k} = k d$ is a dimensionless wave-vector. The reflection matrices implicitly depend on $T$ through the temperature-dependent longitudinal and Hall conductivities. 

In the limit of large distances or high temperatures (the so-called classical limit), the $n=0$ term dominates the whole Matsubara summation, and one needs to evaluate the behavior of the reflection coefficients, and hence the conductivities, for $n=0$. To this end, we perform a low-frequency expansion of the finite-temperature conductivities in Eq.(\ref{sigma_T}), as was done for $T=0$ in the main text, and the corresponding $\tilde{B}^{xx,xy}_i$ coefficients now depend on temperature. We separately study the cases with and without dissipation at finite temperature. 
When $\Gamma \neq 0$, in the limit $\xi \rightarrow 0$ we get
$\sigma_{xx} \approx (\alpha c /2 \pi) \tilde{B}^{xx}_0(T)$, $\sigma_{xy} \approx (\alpha c /2 \pi) \tilde{B}^{xy}_0(T)$, and $R_{\rm sp}(0,\tilde{k}/d)=R_{\rm ps}(0,\tilde{k}/d) \approx 0$, $R_{\rm ss}(0,\tilde{k}/d)\approx 0$, and $R_{\rm pp}(0,\tilde{k}/d) \approx 1$. 
Hence, the layer behaves as a perfectly reflecting interface for p-polarized waves. Note that this is true irrespective of the point $(E_z,\Lambda)$ in phase space and the value of the chemical potential.
The corresponding energy is
\begin{myeqnarray}
\mathcal{E}^{\Gamma \neq 0}_{n=0}(T)=  \frac{k_{\rm B} T}{4 \pi d^2} \int_0^{\infty} d\tilde{k} \; \tilde{k}  
\log (1 - e^{-2 \tilde{k}}) = - \frac{\zeta(3)}{16 \pi} \; \frac{k_{\rm B} T}{d^2},
\end{myeqnarray}
where $\zeta(x)$ is the zeta function. 
When $\Gamma=0$,  in the limit $\xi \rightarrow 0$ we get 
$\sigma_{xx} \approx (\alpha c/2 \pi) \tilde{B}^{xx}_{-1}(T)/\xi$, $\sigma_{xy} \approx (\alpha c/2 \pi) \tilde{B}^{xy}_{0}(T)$, and the reflection coefficients  $R_{\rm sp}(0,\tilde{k}/d)=R_{\rm ps}(0,\tilde{k}/d) \approx 0$, 
$R_{\rm pp}(0,\tilde{k}/d) \approx 1$, and $R_{\rm ss}(0,\tilde{k}/d) \approx - \kappa /(\kappa + \tilde{k})$ where $\kappa=\alpha \tilde{B}^{xx}_{-1}(T) d/c$. In contrast to the dissipative case, $R_{\rm ss}(0,\tilde{k}/d)$ is non-zero and its value is a function of the particular phase of the layer.
The corresponding energy is
\begin{myeqnarray}
\mathcal{E}^{\Gamma=0}_{n=0}(T)=  \mathcal{E}^{\Gamma \neq 0}_{n=0}(T)
+ \frac{k_{\rm B} T}{4 \pi d^2} \int_0^{\infty} d\tilde{k} \; \tilde{k}  
\log \left(1 - \frac{\kappa^2}{(\kappa+\tilde{k})^2} e^{-2 \tilde{k}} \right). 
\end{myeqnarray}
Note that  $\mathcal{E}^{\Gamma=0}_{n=0}(T) \approx 2  \mathcal{E}^{\Gamma \neq 0}_{n=0}(T)$ when $\kappa \gg 1$, and $\mathcal{E}^{\Gamma=0}_{n=0}(T) \approx \mathcal{E}^{\Gamma \neq 0}_{n=0}(T)$ when $\kappa \ll 1$.

\end{widetext}


\begin{thebibliography}{10}

\bibitem{Parsegian_BookVdW}
Parsegian, V.A. {\em {Van der Waals Forces: A Handbook for Biologists, Chemists, Engineers, and Physicists}}.
\newblock (Cambridge University Press, 2005).

\bibitem{Casimir_Polder}
Casimir, H.B.G., Polder D. {The Influence of retardation on the London-van der Waals forces}.
\newblock {\em Phys. Rev.} {\bf 73}, 360-372 (1948).

\bibitem{Casimir_PMPlates}
Casimir, H.B.G. {On the attraction between two perfectly conducting plates}.
\newblock {\em Proceedings Kon. Nederland. Akad. Wetensch.} {\bf 51}, 793-795 (1948).

\bibitem{RPM_Galina_Klimchitskaya}
Klimchitskaya, G..L, Mohideen, U., Mostepanenko, V.M. {The Casimir force between real materials: Experiment and theory}.
\newblock {\em Rev. Mod. Phys.} {\bf 81}, 1827-1885 (2009).

\bibitem{Rodriguez-2011}
Rodriguez, A.W., Capasso, F., Johnson, S.G. {The Casimir effect in microstructured geometries.}
\newblock {\em Nat. Photon.} {\bf 5}, 211-221 (2011).

\bibitem{Dalvit:book}
Dalvit. D.A.R., Milonni, P., Roberts, D., Rosa, F.S.S. {\em {Casimir Physics (Lecture Notes)}}.
\newblock (Springer-Verlag, 2011).

\bibitem{Woods-RevModPhys}
Woods, L.M., Dalvit, D.A.R., Tkatchenko, A., Rodriguez-Lopez, P., Rodriguez, A.W., Podgornik, R. {A materials perspective on Casimir and van der Waals interactions}.
\newblock {\em Rev. Mod. Phys.} {\bf 88}, 045003 (2016).

\bibitem{Novoselov666_Graphene}
Novoselov, K.S., Geim, A.K., Morozov, S.V., Jiang, D., Zhang, Y., Dubonos, S.V., Grigorieva, I.V., Firsov, A.A. {Electric field effect in atomically thin carbon films}.
\newblock {\em Science} {\bf 306}, 666-669 (2004).

\bibitem{PhysRevLett.96.073201}
Dobson, J.F., White, A., Rubio, A.  {Asymptotics of the dispersion interaction: Analytic benchmarks for van der Waals energy functionals}.
\newblock {\em Phys. Rev. Lett.} {\bf 96}, 073201 (2006).

\bibitem{PhysRevB.80.245424}
G{\'o}mez-Santos, G. {Thermal van der Waals interaction between graphene layers}.
\newblock {\em Phys. Rev. B} {\bf 80}, 245424 (2009).

\bibitem{Svetovoy}
Svetovoy, V., Moktadir, Z., Elwenspoek, M., Mizuta, H. {Tailoring the thermal Casimir force with graphene}
\newblock{\em EPL} {\bf 96}, 14006 (2011).

\bibitem{PhysRevB.82.155459}
Drosdoff, D., Woods, L.M. {Casimir forces and graphene sheets}.
\newblock {\em Phys. Rev. B} {\bf 82}, 155459 (2010).

\bibitem{PhysRevB.84.155407}
Sarabadani, J., Naji, A., Asgari, R., Podgornik, R. (2011) {Many-body effects in the van der Waals: Casimir interaction between graphene layers}.
\newblock {\em Phys. Rev. B} {\bf 84}, 155407 (2011).

\bibitem{PhysRevB.89.125407}
Klimchitskaya, G.L., Mostepanenko, V.M., Sernelius, B.E. {Two approaches for describing the Casimir interaction in graphene: Density-density correlation function versus polarization tensor}.
\newblock {\em Phys. Rev. B} {\bf 89}, 125407 (2014).

\bibitem{Quantized_Casimir_Force}
Tse, W.K., MacDonald, A.H. {Quantized Casimir force}.
\newblock {\em Phys. Rev. Lett.} {\bf 109}, 236806 (2012).

\bibitem{VGobre-2014}
Gobre, V.V., Tkatchenko, A. Scaling laws for van der Waals interactions in nanostructured materials.
\newblock {\em Nat. Comm.} {\bf 4}, 2341 (2014).

\bibitem{Tsoi-2014}
Tsoi, S., Dev, P., Friedman, A.L., Stine, R., Robinson, J.T., Reinecke, T.L., Sheehan, P.E. {van der Waals screening by single-layer graphene and molybdenum disulfide}.
\newblock {\em ACS Nano} {\bf 8}, 12410-12417 (2014).

\bibitem{Mohideen-2013}
Banishev, A.A., Wen, H., Xu, J. Kawakami, R.K., Klimchitskaya, G.L., Mostepanenko, V.M., Mohideen, U. {Measuring the Casimir force gradient from graphene on a SiO${}_{2}$ substrate}.
\newblock {\em Phys. Rev. B} {\bf 87}, 205433 (2013).

\bibitem{PhysRevLett.108.155501}
Vogt P et~al. {Silicene: Compelling experimental evidence for graphenelike two-dimensional silicon}.
\newblock {\em Phys. Rev. Lett.} {\bf 108}, 155501 (2012).

\bibitem{Davila-2014}
D{\'a}vila, M.E., Xian, L., Cahangirov, S., Rubio, A., Lay, G.L. {Germanene: a novel two-dimensional germanium allotrope akin to graphene and silicene}.
\newblock {\em New J. of Phys.} {\bf 16}, 095002 (2014).

\bibitem{RIS_0}
Zhu, F.-f., Chen, W.-j., Xu, Y., Gao, C.-l., Guan, D.-d., Liu, C.-h., Qian, D., Zhang, S.-C., Jia, J.-f. {Epitaxial growth of two-dimensional stanene}.
\newblock {\em Nat. Mater.} {\bf 14}, 1020-1026 (2015).

\bibitem{PhysRevLett.107.076802}
Liu, C.-C., Feng, W., Yao, Y. {Quantum spin Hall effect in silicene and two-dimensional germanium}.
\newblock {\em Phys. Rev. Lett.} {\bf 107}, 076802 (2011).

\bibitem{Ezawa-2012}
Ezawa, M. {Valley-polarized metals and quantum anomalous Hall effect in silicene}.
\newblock {\em Phys. Rev. Lett.} {\bf 109}, 055502 (2012).

\bibitem{Ezawa-PhysRevB.86.161407}
Ezawa, M. {Spin-valley optical selection rule and strong circular dichroism in silicene}.
\newblock {\em Phys. Rev. B} {\bf 86}, 161407 (2012).

\bibitem{Ezawa2013}
Ezawa, M.  {Photoinduced topological phase transition and a single Dirac-cone state in silicene}.
\newblock {\em Phys. Rev. Lett.} {\bf 110}, 026603 (2013).

\bibitem{PhysRevLett.111.136804}
Xu, Y, Yan, B., Zhang, H.-J., Wang, J., Xu, G., Tang, P., Duan, W., Zhang, S.-C.  {Large-gap quantum spin Hall insulators in tin films}.
\newblock {\em Phys. Rev. Lett.} {\bf 111}, 136804 (2013).

\bibitem{Houssa2016}
Houssa, M., van der Broek, B., Iordanidou, K., Lu, A.K.A., Pourtos, G., Locquet, J.-P., Afanas'ev, V., Stesmans, A.  {Topological to trivial insulating phase transition in stanene}.
\newblock {\em Nano Research} {\bf 9}, 774-778 (2016).

\bibitem{PhysRevB.86.195405}
Stille, L., Tabert, C.J., Nicol, E.J.  {Optical signatures of the tunable band gap and valley-spin coupling in silicene}.
\newblock {\em Phys. Rev. B} {\bf 86}, 195405 (2012).

\bibitem{PhysRevLett.110.197402}
Tabert, C.J., Nicol, E.J.  {Valley-spin polarization in the magneto-optical response of silicene and other similar 2d crystals}.
\newblock {\em Phys. Rev. Lett.} {\bf 110}, 197402 (2013).

\bibitem{PhysRevB.87.235426}
Tabert, C.J., Nicol, E.J.  {AC/DC spin and valley Hall effects in silicene and germanene}.
\newblock {\em Phys. Rev. B} {\bf 87}, 235426 (2013).

\bibitem{PhysRevB.88.045442}
Xiao, X., Wen, W. {Optical conductivities and signatures of topological insulators with hexagonal warping}.
\newblock {\em Phys. Rev. B} {\bf 88}, 045442 (2013).

\bibitem{VdW_heterostructures}
Geim, A.K., Grigorieva, I.V.  {Van der Waals heterostructures}.
\newblock {\em Nature} {\bf 499}, 419-425 (2013).

\bibitem{Lin-vdWsolids-2014}
Lin, Y.-C., Lu, N., Perea-Lopez, N., Li, J., Lin, Z., Peng, X., Lee, C.H., Sun, C., Calderin, L., Browning, P.N., Bresnehan, M.S., Kim, M.J., Mayer, T.S., Terrones, M., Robinson, J.A.  {Direct synthesis of van der Waals solids}.
\newblock {\em ACS Nano} {\bf 8}, 3715-3723 (2014).

\bibitem{Terrones-2014}
Terrones, H., Del Corro, E., Feng, S., Poumirol, J.M., Rhodes, D., Smirnov, D., Pradhan, N.R., Lin, Z., Nguyen, M.A.T., Elias, A.L., Mallouk, T.E., Balicas, L., Pimenta, M.A., Terrones, M. {New first order Raman-active modes in few layered transition metal dichalcogenides}.
\newblock {\em Sc. Rep.} {\bf 4}, 4215 (2014).

\bibitem{Le-2016}
Le, N.B., Huan, T.D., Woods, L.M.  {Interlayer interactions in van der Waals heterostructures: Electron and phonon properties}.
\newblock {\em ACS Appl. Mater. \& Interfaces} {\bf 8}, 6286-6292 (2016).


\bibitem{PhysRevB.84.195430}
Liu, C.C., Jiang, H., Yao, Y.  {Low-energy effective Hamiltonian involving spin-orbit coupling in silicene and two-dimensional germanium and tin}.
\newblock {\em Phys. Rev. B} {\bf 84}, 195430 (2011).

\bibitem{Floquet_CI_Graphene}
G{\'o}mez-Le{\'o}n, {\'A}., Delplace, P., Platero, G.  {Engineering anomalous quantum Hall plateaus and antichiral states with ac fields}.
\newblock {\em Phys. Rev. B} {\bf 89}, 205408 (2014).

\bibitem{PhysRevLett.112.156801}
Grushin, A.G., G{\'o}mez-Le{\'o}n, {\'A}., Neupert, T. {Floquet fractional Chern insulators}.
\newblock {\em Phys. Rev. Lett.} {\bf 112}, 156801 (2014).

\bibitem{Ezawa-2015}
Ezawa, M. {Monolayer topological insulators: Silicene, germanene, and stanene}.
\newblock {\em J. Phys. Soc. Jap.} {\bf 84}, 121003 (2015).

\bibitem{Oka-2009}
Oka, T., Hideo, A. {Photovoltaic Hall effect in graphene}.
\newblock {\em Phys. Rev. B} {\bf 79}, 081406 (2009).

\bibitem{Kubo-I-1957}
Kubo, R.  {Statistical-mechanical theory of irreversible processes. I. General theory and simple applications to magnetic and conduction problems}.
\newblock {\em J. Phys. Soc. Jap.} {\bf 12}, 570-586 (1957).

\bibitem{Kubo-II-1957}
Kubo, R., Yokota, M., Nakajima, S.  {Statistical-mechanical theory of irreversible processes. II. Response to thermal disturbance}.
\newblock {\em J. Phys. Soc. Jap.} {\bf 12}, 1203-1211 (1957).

\bibitem{RevModPhys.81.109}
Castro Neto, A.H., Guinea, F., Peres, N.M.R., Novoselov, K.S., Geim, A.K.  {The electronic properties of graphene}.
\newblock {\em Rev. Mod. Phys.} {\bf 81}, 109-162 (2009).

\bibitem{Laurent2012}
Laurent, J., Sellier, H., Huant, S., Chevrier, J. {Casimir force measurements in Au-Au and Au-Si cavities at low temperature}.
\newblock{\em Phys. Rev. B} {\bf 85}, 035426 (2012).

\bibitem{Kane-2009}
Kane, C.L., Mele, E.J. {Quantum Hall effect in graphene}.
\newblock {\em Phys. Lett. Lett.} {\bf 95}, 226801 (2009).

\bibitem{CasimirCI}
Rodriguez-Lopez, P., Grushin, A.G. {Repulsive Casimir effect with Chern insulators}.
\newblock {\em Phys. Rev. Lett.} {\bf 112}, 056804 (2014).

\end{thebibliography}

\begin{thebibliography}{10}

\bibitem{Giuliani}
Giuliane, G. F., Vignale, G. {\em {Quantum Theory of the Electron Liquid}}.
\newblock (Cambridge University Press, 2005).

\bibitem{Maldague}
Maldague, P. F. Many-body corrections to the polarizability of the two-dimensional electron gas. {\it Surface Science} {\bf 73}, 296-302 (1978).

\end{thebibliography}
\end{document}